\documentstyle[twoside,12pt,epsfig]{article}
\def\greaterthansquiggle{\raise.3ex\hbox{$>$\kern-.75em\lower1ex\hbox{$\sim$}}}
\def\lessthansquiggle{\raise.3ex\hbox{$<$\kern-.75em\lower1ex\hbox{$\sim$}}}
\newcommand{\beq}{\begin{equation}}
\newcommand{\eeq}{\end{equation}}
\newcommand{\beqa}{\begin{eqnarray}}
\newcommand{\eeqa}{\end{eqnarray}}
\newcommand{\ba}{\begin{array}}
\newcommand{\ea}{\end{array}}

\newcommand{\grts}{\greaterthansquiggle}
\newcommand{\lets}{\lessthansquiggle}

\newcommand{\ra}{\rightarrow}

\newcommand{\wt}{\widetilde}

\newcommand{\fb}{{\rm f\/b}}

\newcommand{\recht} {\begin{picture}(7,7)
                      \linethickness{1.7mm}
                      \put(2,1){\line(0,1){5}}
                      \thinlines
                     \end{picture}
                    } 
\newcommand{\rechtl} {\begin{picture}(7,7)
                      \put(2,1){\framebox(5,5){}}
                     \end{picture}
                    }

\def\noi             {\noindent}

\def\x               {\chi}
\def\ee              {$e^+ e^-$}
\def\eeto            {e^+ e^- \to}

\def\st              {\wt t}

\def\stst            {\st_1\,\bar{\st}_1}

\def\sb              {\wt b}

\def\sbsb            {\sb_1\,\bar{\sb}_1}

\def\stau            {\wt \tau}

\def\staustau        {\stau_1\,\bar{\stau}_1}

\def\sf              {\wt f}

\def\chp             {\wt \x^+}
\def\chm             {\wt \x^-}
\def\nt              {\wt \x^0}
\newcommand{\mst}[1]   {M_{\st_{#1}}}
\newcommand{\msb}[1]   {M_{\sb_{#1}}}
\newcommand{\mstau}[1] {M_{\stau_{#1}}}

\newcommand{\mch}[1]   {M_{\wt \x^\pm_{#1}}}
\newcommand{\mnt}[1]   {M_{\wt \x^0_{#1}}}

\def\ptmiss           {p\llap/_{T}}

\newcommand{\etal }   {{\em et~al.}}

\newcommand{\chichi}{\mbox{$\tilde{\chi}^0_1c\tilde{\chi}^0_1\bar{c}$}}
\newcommand{\chipchip}{\mbox{$\tilde{\chi}^+_1b\tilde{\chi}^-_1\bar{b}$}}

\newcommand{\chiz}  {\mbox{$\tilde{\chi}^0_1$}}
\newcommand{\chip}  {\mbox{$\tilde{\chi}^+_1$}}
\newcommand{\chim}  {\mbox{$\tilde{\chi}^-_1$}}

\newcommand{\cc}      {\mbox{$c \overline{c}$}}
\newcommand{\gev}      [1]{\mbox{$\mathrm{  #1 \ GeV      }$}}
\newcommand{\bb}      {\mbox{$b \overline{b}$}}
\newcommand{\abs}[1]{\mbox{$ |#1|        $}}      % absolute
\newcommand{\tautau}  {\mbox{$\rm \tau^+ \tau^- $}}
\newcommand{\mm}      {\mbox{$\rm \mu^+ \mu^- $}}
\newcommand{\qq}      {\mbox{$q \overline{q}$}}
\newcommand{\ZZ}        {\mbox{$Z^0 Z^0$}}
\newcommand{\WW}        {\mbox{$W^+W^-$}}
\newcommand{\lumifb}   [1]{\mbox{$\cal L = \invfb{#1}$}}
\newcommand{\invfb}    [1]{\mbox{$\mathrm{  #1 \  fb^{-1}}$}}
\newcommand{\sqrts} {\mbox{$\sqrt{s}$}}

\normalsize
\begin{document}
\bibliographystyle{plain}

%------------------------------------------------------------------------
\begin{titlepage}
%------------------------------------------------------------------------

\begin{flushright}
UWThPh-1996-20\\
HEPHY-PUB 644/96\\
hep-ph/9604221\\
\vspace{0.2cm}
March, 1996
\end{flushright}
\vspace*{1.5cm}

\begin{center}

\begin{LARGE}
  Search of Stop, Sbottom, and Stau at an \\
  $e^+e^-$ Linear Collider with $\sqrt{s}=0.5 - 2$~TeV\footnote{
Contribution to the 
Workshop on Physics with $e^+ e^-$ Linear Colliders, Annecy -- Gran Sasso --
Hamburg, 1995, ed. P.~Zerwas}\\
\end{LARGE}

\vspace{2cm}

\begin{large} 
A. Bartl$^{\small 1}$, 
H. Eberl$^{\small 2}$, 
S. Kraml$^{\small 2}$,\\[1mm]
W. Majerotto$^{\small 2}$, 
W. Porod$^{\small 1}$,
A. Sopczak$^{\small 3}\footnote{c/o PPE Division, CERN}$ \\[8mm]
\end{large}

{\em (1) Institut f\"ur Theoretische Physik, Universit\"at Wien,
           A-1090 Vienna, Austria } \\[1mm]
{\em (2) Institut f\"ur Hochenergiephysik, \"Osterreichische Akademie der 
           Wissenschaften, A-1050 Vienna, Austria } \\[1mm]
{\em (3) DESY-Zeuthen, D-15738, Zeuthen, Germany} \\ 

\end{center}

\vfill
\begin{abstract}

We discuss pair production and decays of stops, sbottoms, and
staus in $e^+e^-$ annihilation in the energy range $\sqrt{s} =
500$~ GeV to $2$~TeV. We present numerical
predictions within the Minimal Supersymmetric Standard
Model for cross sections and decay rates.
We study the stop discovery potential for
$\sqrt{s} = 500$~GeV and $10~\fb^{-1}$ integrated luminosity with
full statistics background simulation.
\end{abstract}

\end{titlepage}

\setcounter{page}{2}

%------------------------------------------------------------------------
\section{Introduction}
%------------------------------------------------------------------------

In the experimental search for supersymmetry (SUSY) particular
attention is paid to those particles which are expected to be
relatively light. The scalar top quark,
the SUSY partner of the top quark, may be the lightest squark,
and may even be the lightest visible SUSY particle (LVSP) 
\cite{Ellis,Altarelli}.
The stop can be light for two reasons: (i) Due to the large top Yukawa
terms in the renormalization group equations, the scalar mass parameters of the
stop can be much smaller than the corresponding parameters of the first and
second generation squarks \cite{Boer,Drees}. (ii) The
off-diagonal elements of the mass mixing matrix
of the stop can be large, and this leads to strong
$\wt t_L - \wt t_R$ mixing.
If the parameter $\tan \beta$ is large enough ($\tan \beta \grts 10$)
the scalar bottom quark \cite{Bartl94} or the scalar tau lepton could
also be relatively
light and even be the LVSP. The existence of a
relatively light stop would have many interesting
phenomenological implications. A light stop would significantly
influence the branching ratios of the decays
$Z^0 \ra b \bar b$, $t \ra bW$, $b \ra s \gamma$ and some other
physical observables (see, e.g. \cite{Kon}).

In this contribution we shall present results
for the production of stops, sbottoms, and staus in
$e^+e^-$ annihilation at energies between $\sqrt{s} = 500$~GeV
and $2$~TeV and details on signal selection and background rejection for stop
production at $\sqrt{s} =$ 500 GeV and \mbox{$\cal{L} = \invfb{10}$}.
The production cross sections and the decay
rates, and thus the discovery reach of these sfermions show a
distinct dependence on
the L--R mixing angles. The most important decay modes of these
sfermions are those into fermions and neutralinos or charginos.

Our framework is the Minimal Supersymmetric Standard
Model (MSSM) \cite{Haber} which contains the Standard Model (SM)
particles, sleptons, $\wt \ell^\pm$, $\wt \nu_\ell$,
squarks, $\wt q$, gluinos $\wt g$, two pairs of charginos, $\wt
\chi^\pm_i$, $i = 1, 2$, four neutralinos, $\wt \chi^0_i$,
$i = 1,\ldots,4$, and five Higgs particles, $h^0$, $H^0$, $A^0$,
$H^\pm$ \cite{Gunion}.
The phenomenology of stops, sbottoms, staus, and their
decay products is determined by the following parameters:
$M$ and $M'$, the (soft breaking) $SU(2)$ and $U(1)$ gaugino masses,
$\mu$, the higgsino mass parameter,
$\tan\beta = v_{2}/v_{1}$ (where $v_1$ and $v_2$ are the vacuum expectation
values of the neutral members of the two Higgs doublets), and
$M_{\wt L}$,  $M_{\wt E}$, $M_{\wt Q}$,  $M_{\wt U}$, $M_{\wt D}$,
$A_{\tau}$, $A_t$, $A_b$, which are
soft--breaking parameters entering the mass mixing matrices of
the stau, stop, and sbottom systems. We assume the GUT relations
$M'/M = \frac{5}{3} \tan^2\Theta_W \approx 0.5$, and
$m_{\wt g}/M = \alpha_s/\alpha_2 \approx 3$, where $m_{\wt g}$ is
the gluino mass. Furthermore, we assume that the
$\nt_1$ is the lightest SUSY particle (LSP).

The lower model independent mass bound for stops obtained at LEP
is 45 GeV \cite{Grivaz,Opal94}.
Stronger limits up to 55 GeV are reported from the data taking at LEP at
130--140 GeV \cite{Nowak}.
The D0 experiment at the TEVATRON excludes the
mass range $40$~GeV $\lets~M_{\wt t}~\lets 100$~GeV for the
stop, if the mass difference $M_{\wt t} - m_{\wt \chi^0_1}~\grts~30$~GeV
\cite{D0}.

In Section~2 we shortly review the basic facts about L--R mixing
of stops, sbottoms, and staus, and present our numerical results for the
production cross sections for unpolarized beams as well as for
polarized $e^-$ beams.
In Section~3 we describe the decays of stops, sbottoms, and
staus and present numerical results for the important branching
ratios. We also list the signatures which are expected to be relevant
at $\sqrt{s} = 500$ GeV. In Section~4 we describe an event generator for
$\stst$ production and decay.
In Section~5 experimental sensitivities are determined based on
Monte Carlo simulations. Section~6 contains a summery.

%------------------------------------------------------------------------
\section{Cross Sections for Pair Production of
         Stops, Sbottoms, and Staus}
%------------------------------------------------------------------------

The SUSY partners of the SM fermions with left and right
helicity are the left and right sfermions. In the case
of the stop, sbottom and stau the left and right states are in
general mixed. In the ($\tilde{f}_L,\tilde{f}_R$) basis
the mass matrix is \cite{Ellis,Gunion}
\beqa
M^2_{\tilde{f}} = \left( \begin{array}{cc}
                        M^2_{\tilde{f}_L} & a_f m_f \\
                        a_f m_f & M^2_{\tilde{f}_R}
                       \end{array}
                 \right)
\eeqa
with
\beqa
  &&\hspace{-11mm} M^2_{\wt f_L} = M^2_{\wt F} +
  m_Z^2 \cos 2\beta(T^3_f - e_f \sin^2\Theta_W) + m^2_f , \\
  &&\hspace{-11mm} M^2_{\wt f_R} = M^2_{\wt F'} + 
  e_f m_Z^2 \cos 2\beta \sin^2\Theta_W + m^2_f , \\
  &&\hspace{-11mm} m_t a_t \equiv m_t(A_t - \mu \cot \beta), \hspace{2mm} 
  m_b a_b \equiv m_b(A_b - \mu \tan \beta), \hspace{2mm} 
  m_{\tau} a_{\tau} \equiv m_{\tau}(A_{\tau} - \mu \tan\beta),
\eeqa
where $e_f$ and $T^3_f$ are the charge and the third component
of weak isospin of the sfermion $\wt f$, $M_{\wt F}=M_{\wt
Q}$ for $\wt f_L = \wt t_L, \wt b_L$, $M_{\wt F}=M_{\wt L}$ for
$\wt f_L = \wt \tau_L$, $M_{\wt F'} = M_{\wt U}, M_{\wt D},
M_{\wt E}$ for $\wt f_R = \wt t_R, \wt b_R, \wt \tau_R$,
respectively, and $m_f$ is the mass of the corresponding
fermion.
Evidently, there can
be strong $\wt t_L$-$\wt t_R$ mixing due to the large top quark
mass. Similarly, for sbottoms and staus L--R mixing is non-negligible if
$\tan \beta \;\grts\; 10$. The mass eigenvalues for the sfermion
$\wt f = \wt t, \wt b, \wt \tau$  are
\beq
  M^2_{\wt f_{1,2}} = \frac{1}{2} \left(M^2_{\wt f_L} + M^2_{\wt f_R}
  \mp \sqrt{(M^2_{\wt f_L} - M^2_{\wt f_R})^2 +
  4m_f^2 a^2_f} \,\right)\\
\eeq
where $\wt t_1$,  $\wt b_1$ and $\wt \tau_1$ denote the lighter eigenstates.

It is well known that the cross section for
$e^+ e^- \ra \wt t_1 \bar{\wt t_1}$ depends
on the stop--mixing parameters. In particular the $Z^0\wt t_1 \bar{\wt
t_1}$ coupling vanishes for the mixing angle 
$\Theta_{\wt t} = 0.98$ \cite{Drees90}.
The cross sections for
 $e^+ e^- \ra \wt b_1 \bar{\wt b_1}$ and
 $e^+ e^- \ra \wt \tau_1 \bar{\wt \tau_1}$ also show a
characteristic dependence on their mixing angles.
The $Z^0\wt b_1 \bar{\wt b_1}$ coupling
vanishes at $\Theta_{\wt b} = 1.17$, and the
$Z^0\wt \tau_1 \bar{\wt \tau_1}$ coupling vanishes at
$\Theta_{\wt \tau} = 0.82$. The interference between
the $\gamma $ and $Z^0$ exchange contributions leads to
characteristic minima of the cross sections 
for $e^+ e^- \ra \wt f_1 \bar{\wt f_1}$ which occur
at specific values of the mixing angles $\theta_{\wt f}$.
They are given by
\beq
  \cos^2\Theta_{\wt f}|_{min} = \frac{e_f}{T^3_f} \sin^2\Theta_W
  [1+(1-s/m^2_Z)  F(\sin^2 \Theta_W)]\, .
\eeq

\noi The function $F(\sin^2 \Theta_W)$
depends on the polarization of the $e^-$ beam and is given by
$F(\sin^2 \Theta_W) = \cos^2 \Theta_W (L_e + R_e)/(L_e^2 +
R_e^2) \approx -0.22$, $F(\sin^2 \Theta_W) = \cos^2 \Theta_W
/L_e \approx -2.9$, and  $F(\sin^2 \Theta_W) = \cos^2 \Theta_W
/R_e \approx 3.3$, for unpolarized, left and right polarized
$e^-$ beams, respectively, where $L_e = - \frac{1}{2} + \sin^2
\Theta_W$ and $R_e = \sin^2 \Theta_W$. For polarized $e^-$
beams the dependence on the mixing angles is much more
pronounced than for unpolarized beams.
The corresponding minima of the cross sections of
$\eeto\sf_{2}\bar{\sf_{2}}$ occur at $1 - \cos^{2}\Theta_{\sf}|_{min}$.

In the calculations of the cross sections
we have used the tree level formulae of \cite{Drees90,Hikasa,Bartl96}.
We have also included SUSY QCD corrections taking the formulae of
\cite{Eberl} (see also \cite{Drees90} and \cite{Beenaker}) and corrections due
to initial state radiation \cite{Peskin}.

In Fig. 1a we show contour lines  of the total cross section
$e^+ e^- \ra \wt t_1 \bar{\wt t_1}$ in the
$M_{\wt t_1} - \cos^2\Theta_{\wt t}$ plane for $\sqrt{s} =
500$~GeV and unpolarized beams. For $M_{\wt t_1} \simeq$ 100~GeV
this cross section can reach 220~fb.
A substantial dependence on $\cos^2\Theta_{\wt t}$  can be
seen for $M_{\wt t_1} \lets 150$~GeV. 
In Fig.~1b we show the $\cos^2\Theta_{\wt t}$ dependence of the cross section
$e^+ e^- \ra \wt t_1 \bar{\wt t_1}$  for left and right
polarized and unpolarized $e^-$ beams for $\sqrt{s} =
500$~GeV and  $M_{\wt t_1} = 200$~GeV. 
The polarization asymmetry depends quite
strongly on the mixing angle. Therefore,
experiments with polarized $e^-$ beams would be necessary
for a precise determination of the mixing
angle $\Theta_{\wt t}$.
The determination of the stop masses and mixing angle gives information on
the basic SUSY parameters $M_{\wt Q}$,  $M_{\wt U}$ and $A_t$.
This is discussed in \cite{Bartl94a}.

Similarly, Fig. 2 is a contour plot of the total cross section
of $e^+ e^- \ra \wt t_2 \bar{\wt t_2}$ in the $M_{\wt t_2} -
\cos^2\Theta_{\wt t}$ plane at $\sqrt{s} = 2$~TeV, for left and
right polarized $e^-$ beams. Here we observe a
strong dependence on the stop mixing angle.  
For $M_{\wt t_2} \simeq$ 900 GeV the cross section at this energy is
about 1 fb.

In Fig. 3 we show the cross section for
$e^+ e^- \ra \wt t_1 \bar{\wt t_2}  +  \bar{\wt t_1} \wt t_2$ 
 at $\sqrt{s}=1$~TeV as
a function of $M_{\wt t_1}$, for various values of $M_{\wt t_2}$.
Here we have fixed the mixing angle $\cos^2 \Theta_{\wt t}=0.5$
where the cross section has its maximum. For other values of the
mixing angle this cross section scales as 
$\sin^2\Theta_{\wt t}\,\cos^2\Theta_{\wt t}$. 
For $M_{\wt t_1} \simeq$ 100~GeV and $M_{\wt t_2} \simeq$ 200~GeV
this cross section can reach $20~\fb$.
Note that $e^+ e^- \ra \wt b_1 \bar{\wt b_2}$ has the same cross
section as $e^+ e^- \ra \wt t_1 \bar{\wt t_2}$,
if the masses and the mixing angles are the same
(neglecting corrections due to gluino exchange).
The cross section for $e^+ e^- \ra \wt \tau_1 \bar{\wt \tau_2}$
is a factor of approximately $1/3$ smaller because of the
colour factor and the QCD radiative corrections which have to
be included for squark production. Due to the factor  
$\sin^2\Theta_{\wt f}\,\cos^2\Theta_{\wt f}$, the
cross section depends strongly on the mixing angle.

In Fig. 4a we show the
contour plot of the total cross section of 
$e^+ e^- \ra \wt b_1 \bar{\wt b_1}$ in the $M_{\wt b_1} -
\cos^2\Theta_{\wt b}$ plane at $\sqrt{s} = 1$~TeV, for
unpolarized beams. For $M_{\wt b_1} \simeq$ 100~GeV (450~GeV)
this cross section can reach a value of about
50~fb (1~fb).
For $M_{\wt b_1} \lets 300$~GeV the cross section 
depends appreciably on $\cos^2\Theta_{\wt b}$. For polarized
$e^-$  beams we have again a much stronger  $\cos^2\Theta_{\wt b}$
dependence of the cross sections, as shown in Fig. 4b
\noi
\begin{minipage}[t]{73mm}   % --- fig.1a ---
{\setlength{\unitlength}{1mm}
\begin{picture}(73,76)                        
\put(3,4){\mbox{\epsfig{figure=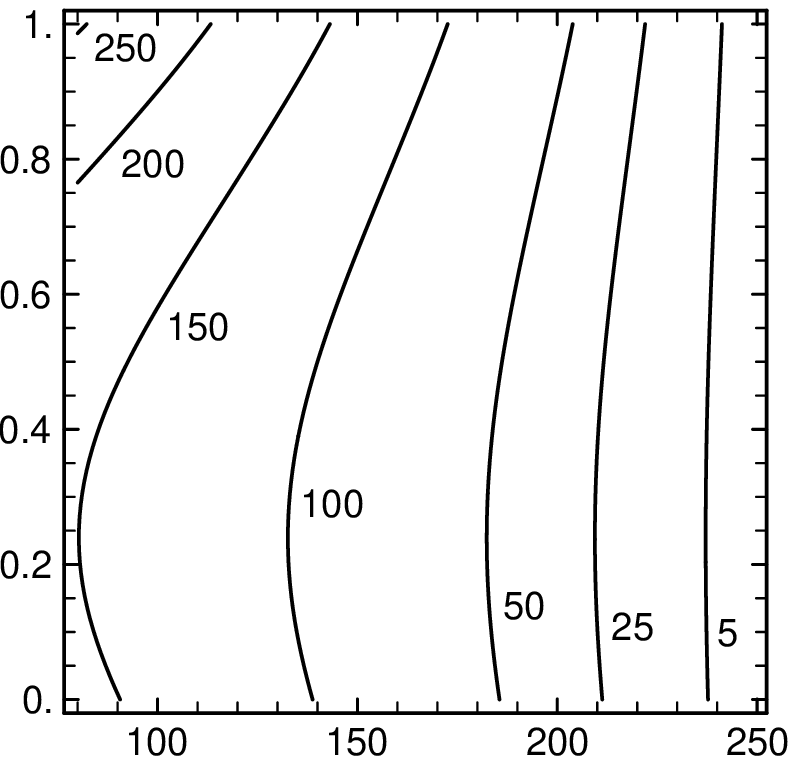,height=6.6cm}}}
\put(69.5,0.3){\makebox(0,0)[br]{{\small $\mst{1}$~[GeV]}}}
\put(3,70){\makebox(0,0)[bl]{{\small $\cos^{2}\Theta_{\st}$}}}
\end{picture}}
\label{fig:st11prod}
\begin{small}
  {\bf Fig.~1a:}~Contour lines for the
  total cross section of $\eeto\stst$ in f\/b at 
  $\sqrt{s} = 500$~GeV 
  as a function of $\cos^{2}\Theta_{\st}$ and $\mst{1}$.
\end{small} 
\end{minipage}
\hspace{3mm}
\begin{minipage}[t]{73mm}   % --- fig.1b ---
{\setlength{\unitlength}{1mm}
\begin{picture}(73,76)                        
\put(3,4){\mbox{\epsfig{figure=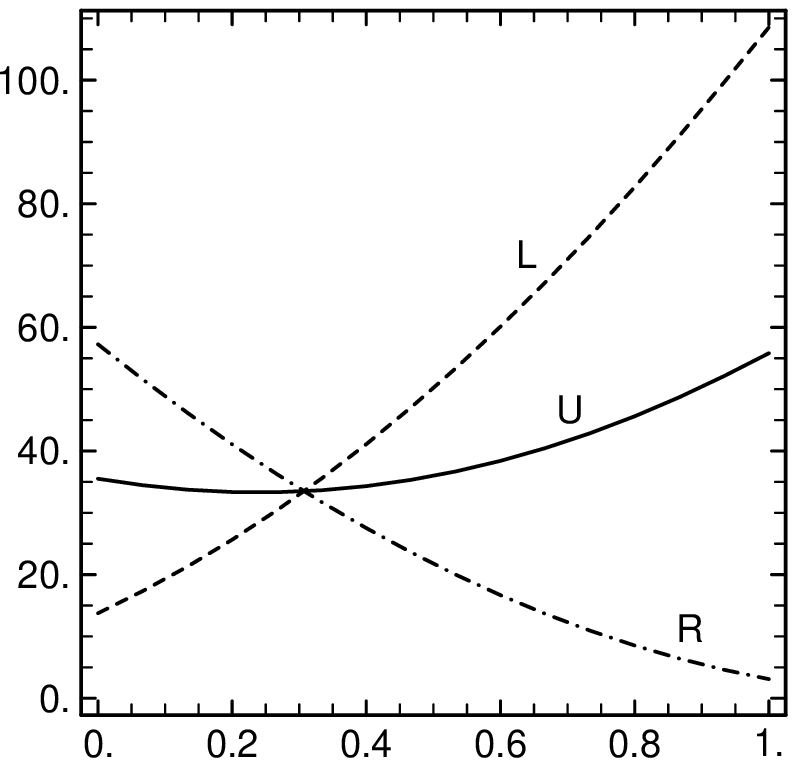,height=6.6cm}}}
\put(69.5,0.5){\makebox(0,0)[br]{{\small $\cos^{2}\Theta_{\st}$}}}
\put(3,70){\makebox(0,0)[bl]{{\small $\sigma(\stst )$~[f\/b]}}}
\end{picture}}
\refstepcounter{figure}   
\label{fig:st11pol}
\begin{small}
  {\bf Fig.~1b:}~Total cross section of $\eeto\stst$ in f\/b
   at $\sqrt{s} = 500$~GeV
   as a function of $\cos^{2}\Theta_{\st}$, for unpolarized (U) as well as
   left (L) and right (R) polarized $e^{-}$ beams and $\mst{1} = 200$~GeV.
\end{small} 
\end{minipage}
\vspace{5mm}

\noi
\begin{minipage}[t]{73mm}   % --- fig.2 ---
{\setlength{\unitlength}{1mm}
\begin{picture}(73,76)                        
%\put(0,0){\framebox(73,76){}}
\put(3,4){\mbox{\epsfig{figure=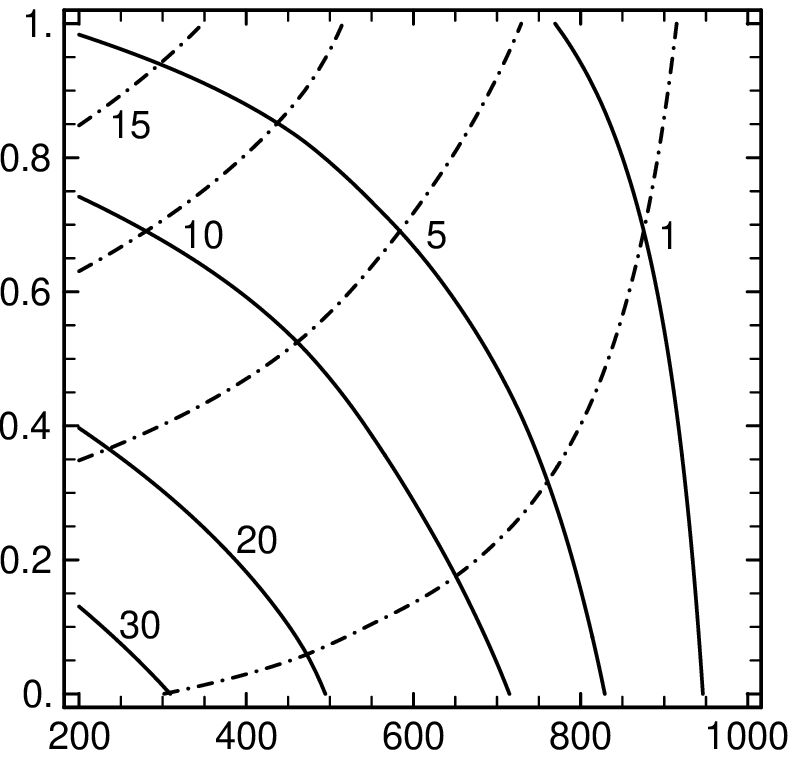,height=6.6cm}}}
\put(69.5,0.3){\makebox(0,0)[br]{{\small $\mst{2}$~[GeV]}}}
\put(3,70){\makebox(0,0)[bl]{{\small $\cos^{2}\Theta_{\st}$}}}
\end{picture}}
\refstepcounter{figure}    
\label{fig:st22prod}
\begin{small}
  {\bf Fig.~2:}~Contour lines for the
  total cross section of $\eeto\st_{2}\bar{\st}_{2}$ in f\/b at 
  $\sqrt{s} = 2$~TeV 
  as a function of $\cos^{2}\Theta_{\st}$ and $\mst{2}$ for left 
  (solid lines) and right (dashdotted lines) polarized $e^{-}$ beams.
\end{small} 
\end{minipage}
\hspace{3mm}
\begin{minipage}[t]{73mm}   % --- fig.3 ---
{\setlength{\unitlength}{1mm}
\begin{picture}(73,76)                        
%\put(0,0){\framebox(73,76){}}
\put(3,4){\mbox{\epsfig{figure=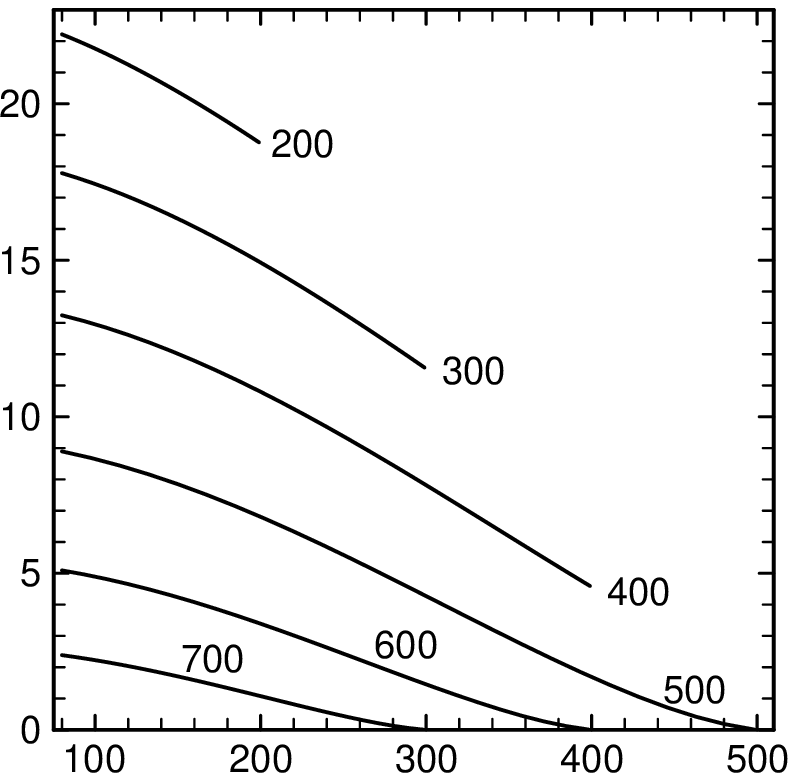,height=6.6cm}}}
\put(69.5,0.3){\makebox(0,0)[br]{{\small $\mst{1}$~[GeV]}}}
\put(3,70){\makebox(0,0)[bl]{{\small $\sigma(\st_{1}\bar{\st}_{2} + 
{\rm c.c.})$~[f\/b]}}}
\end{picture}}
\refstepcounter{figure}
\label{fig:st12prod}
\begin{small}
  {\bf Fig.~3:}~Total cross section of $\eeto\st_{1}\bar{\st}_{2}$ + c.c. 
  in f\/b at $\sqrt{s} = 1$~TeV 
  as a function of $\mst{1}$ for $\cos^{2}\Theta_{\st} = 0.5$ and 
  various masses of $\st_{2}$.
\end{small} 
\end{minipage}
\newpage

\vspace{3mm}
\noi
\begin{minipage}[t]{73mm}   % --- fig.4a ---
{\setlength{\unitlength}{1mm}
\begin{picture}(73,76)                        
%\put(0,0){\framebox(73,76){}}
\put(3,4){\mbox{\epsfig{figure=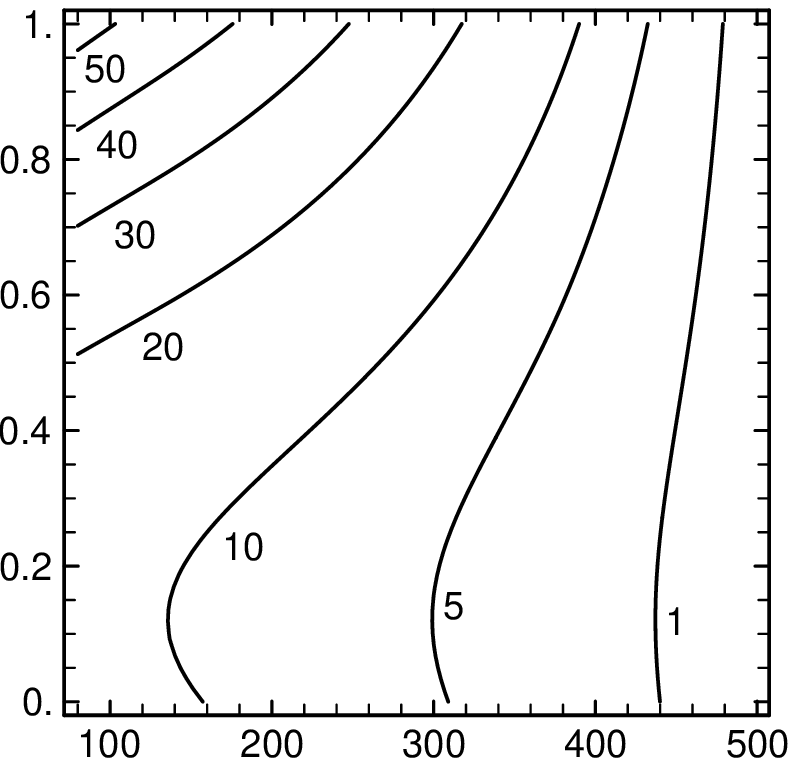,height=6.6cm}}}
\put(69.5,0.3){\makebox(0,0)[br]{{\small $\msb{1}$~[GeV]}}}
\put(3,70){\makebox(0,0)[bl]{{\small $\cos^{2}\Theta_{\sb}$}}}
\end{picture}}
%\refstepcounter{figure}    
\label{fig:sb11prod}
\begin{small}
  {\bf Fig.~4a:}~Contour lines for the
  total cross section of $\eeto\sbsb$ in fb at
  $\sqrt{s} = 1$~TeV 
  as a function of $\cos^{2}\Theta_{\sb}$ and $\msb{1}$.
\end{small} 
\end{minipage}
\hspace{3mm}
\begin{minipage}[t]{73mm}   % --- fig.4b ---
{\setlength{\unitlength}{1mm}
\begin{picture}(73,76)                        
%\put(0,0){\framebox(73,76){}}
\put(3,4){\mbox{\epsfig{figure=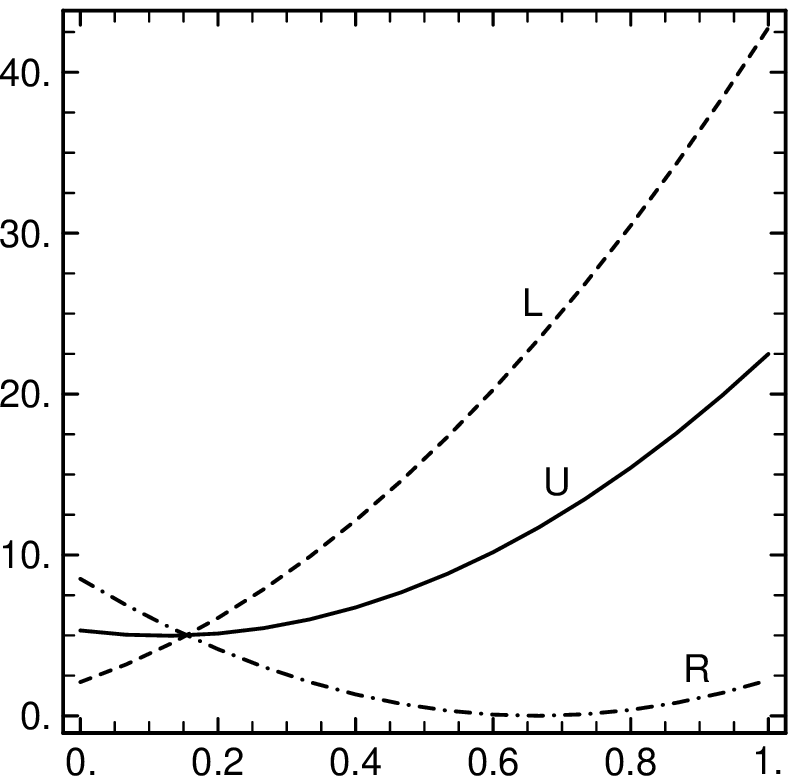,height=6.6cm}}}
\put(69.5,0.3){\makebox(0,0)[br]{{\small $\cos^{2}\Theta_{\sb}$}}}
\put(3,70){\makebox(0,0)[bl]{{\small $\sigma(\sbsb )$~[f\/b]}}}
\end{picture}}
\refstepcounter{figure}   
\label{fig:sb11pol}
\begin{small}
  {\bf Fig.~4b:}~Total cross section of $\eeto\sbsb$ in f\/b 
   at $\sqrt{s} = 1$~TeV
   as a function of $\cos^{2}\Theta_{\sb}$, for unpolarized (U), and
   left (L), and right (R) polarized $e^{-}$ beams for $\msb{1} = 300$~GeV.
\end{small} 
\end{minipage}
\vspace{3mm}

\noi
\begin{minipage}[t]{73mm}   % --- fig.5 ---
{\setlength{\unitlength}{1mm}
\begin{picture}(73,76)
%\put(0,0){\framebox(73,76){}}
\put(3,4){\mbox{\epsfig{figure=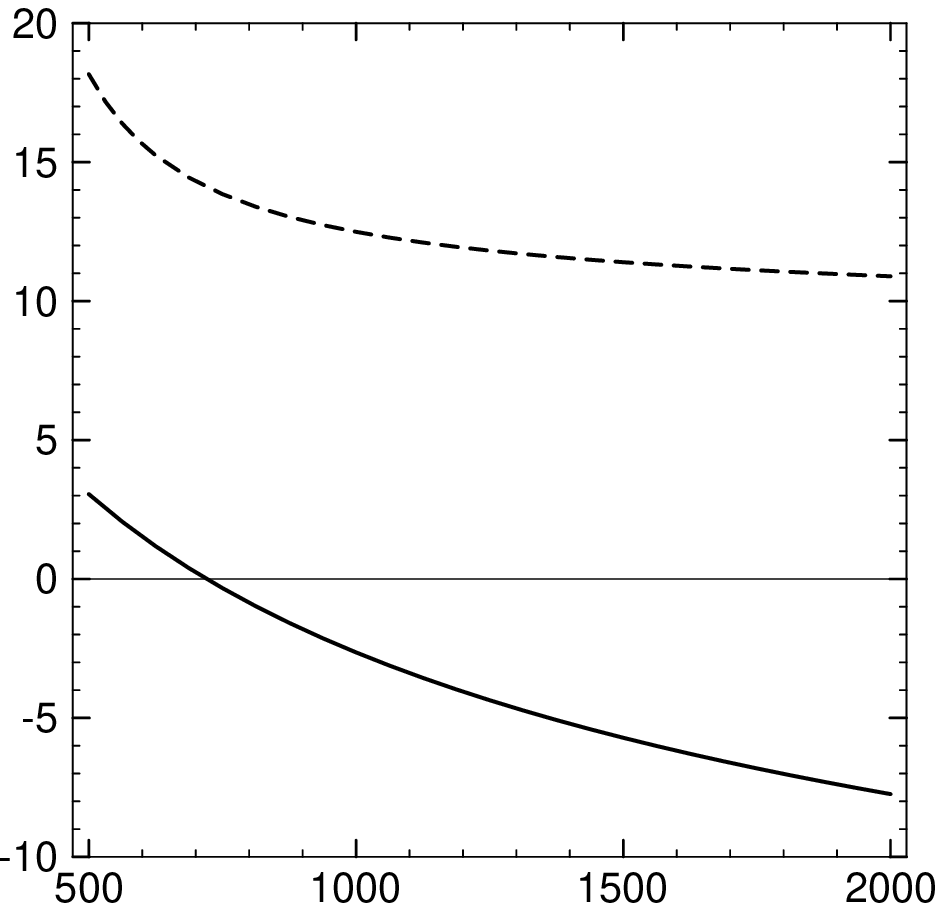,height=6.6cm}}}
\begin{small}
\put(69,0.5){\makebox(0,0)[br]{{\boldmath{$\sqrt{s}$}}~[GeV]}}
\put(0,70){\makebox(0,0)[bl]{\bf{\%}}}
\put(39,38){\makebox(0,0)[b]{\boldmath{$e^+ e^- \to \tilde{t}_1
\bar{\tilde{t}}_1$}}}
\put(16,57){\makebox(0,0)[bl]{$\delta\sigma^g/\sigma^{tree}$}}
\put(50,18){\makebox(0,0)[b]{$\delta\sigma^{\tilde g}/\sigma^{tree}$}}
\end{small}
\end{picture}}
\refstepcounter{figure}    
\label{fig:QCDcorr}
\begin{small}
  {\bf Fig.~5:}~SUSY--QCD corrections $\delta \sigma^g / \sigma^{tree}$
  and $\delta \sigma^{\tilde g} / \sigma^{tree}$ for
  $\eeto\stst$ as a function of $\sqrt{s}$ for $\cos\Theta_{\st} = 0.7,
  M_{\wt t_1} =$ 150 GeV, $M_{\wt t_2} =$ 300 GeV and $M_{\wt g} =$ 300 GeV. 
\end{small} 
\end{minipage}
\hspace{3mm}
\begin{minipage}[t]{73mm}   % --- fig.6 ---
{\setlength{\unitlength}{1mm}
\begin{picture}(73,76)
%\put(0,0){\framebox(73,76){}}
\put(3,4){\mbox{\epsfig{figure=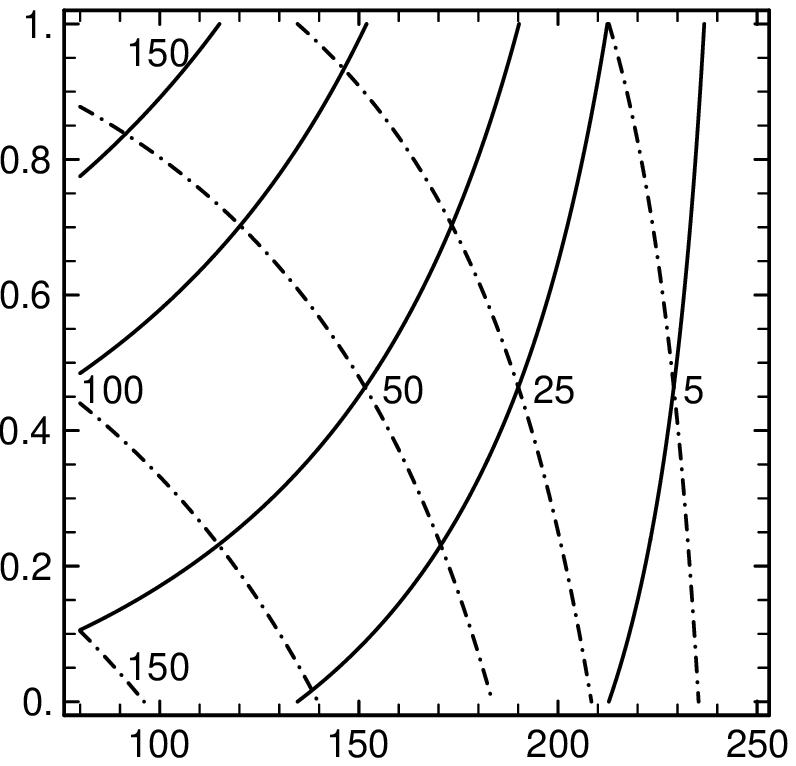,height=6.6cm}}}
\put(69.5,0.5){\makebox(0,0)[br]{{\small $\mstau{1}$~[GeV]}}}
\put(3,70){\makebox(0,0)[bl]{{\small $\cos^{2}\Theta_{\stau}$}}}
\end{picture}}
\refstepcounter{figure}    
\label{fig:sl11prod}
\begin{small}
  {\bf Fig.~6:}~Contour lines for the
  total cross section of $\eeto\staustau$ in f\/b at 
  $\sqrt{s} = 500$~GeV 
  as a function of $\cos^{2}\Theta_{\stau}$ and $\mstau{1}$ for left 
  (solid lines) and right (dashdotted lines) polarized $e^{-}$ beams.
\end{small} 
\end{minipage}
\vspace{3mm}

\noi
for  $M_{\wt b_1} = 300$~GeV and  $\sqrt{s} = 1$~TeV.

The influence of the SUSY QCD corrections as a function of $\sqrt{s}$
is demonstrated in Fig. 5, where we have taken $\cos\Theta_{\st} = 0.7,
M_{\wt t_1} =$ 150 GeV, $M_{\wt t_2} =$ 300 GeV and $m_{\wt g} =$ 300 GeV. 
$\delta \sigma^{g}$ is the conventional QCD correction and
$\delta \sigma^{\tilde g}$ is the correction due to gluino exchange.
Note that at high energies $\delta \sigma^{\tilde g}$ has the opposite sign
of $\delta \sigma^{g}$, and its absolute value is increasing with $\sqrt{s}$.
For a more detailed discussion of SUSY QCD corrections see \cite{Eberl}. 
They increase the cross section values up to 40 \%.
The corrections due to initial state radiation turn out to be 
of the order of 10 \%.

The cross sections for  $e^+ e^- \ra \wt \tau_1 \bar{\wt
\tau_1}$ at $\sqrt{s}=500$~GeV for left and right polarized
$e^-$ beams, as a function of $M_{\wt \tau_1}$ and 
$\cos^2 \Theta_{\wt \tau}$ are shown in Fig. 6. For both beam
polarizations these cross sections can reach values of
approximately $150~\fb$, again exhibiting a strong dependence 
on the mixing angle.

%------------------------------------------------------------------------
\section{Stop, Sbottom, and Stau Decays}
%------------------------------------------------------------------------

The sfermions of the third generation can decay according to
\beqa
  \wt t_i &\ra& t \wt \chi^0_k, \qquad  b \wt \chi^+_k
  \label{stferm} \\
  \wt b_i &\ra& b \wt \chi^0_k, \qquad  t \wt \chi^-_k
  \label{sbferm} \\
  \wt \tau_i &\ra& \tau \wt \chi^0_k, \qquad  \nu_{\tau} \wt \chi^+_k
\eeqa
Due to the Yukawa terms and
because of L--R mixing the decay patterns of stops, sbottoms,
and staus will be different from
those of the sfermions of the first two generations \cite{Bartl91}.
Stops and sbottoms may also decay into gluinos, 
\beq
  \wt t_i \ra t \wt g, \qquad \wt b_i \ra b \wt g,
 \label{sqglu}
\eeq
and if these decays are kinematically allowed, then they are dominant.
Otherwise, the decays (\ref{stferm}), (\ref{sbferm}) 
are the most important ones.
Moreover, in case of strong L--R mixing the splitting
between the two mass eigenstates may be so large that the
following additional decay modes are present \cite{Bartl94}:
$\wt t_2 \ra \wt t_1 \, Z^0 (h^0,\, H^0,\, A^0)$, 
$\wt b_1 \, W^+ (H^+)$,
$\wt b_2 \ra \wt b_1 \, Z^0 (h^0,\, H^0,\, A^0)$,
$\wt t_1 \, W^- (H^-)$.
The transitions $\wt t_1 \ra \wt b_1 \, W^+ (H^+)$
or $\wt b_1 \ra \wt t_1 \, W^- (H^-)$ can occur if the mass difference
is large enough. 

If the $\wt t_1$ is the LVSP and 
$m_{\wt \chi^0_1} + m_b + m_W < M_{\wt t_1} < m_{\wt \chi^0_1} + m_t$,
then the decay $\wt t_1 \ra b \, W^+ \, \wt \chi^0_1$
is important.
If $ M_{\wt t_1} < m_{\wt \chi^0_1} + m_b + m_W$ the higher--order decay
$\wt t_1 \ra c \, \wt \chi^0_1$ dominates \cite{Hikasa}. 
In the parameter domain where $\wt t_1 \ra b \, W^+ \, \wt \chi^0_1$
is possible it is usually more important than $\wt t_1 \ra c \, \wt \chi^0_1$.
If $\wt b_1$ or $\wt \tau_1$ is the LVSP, then it decays
according to $\wt b_1 \ra b \, \wt \chi^0_1$ or 
$\wt \tau_1 \ra \tau \, \wt \chi^0_1$. 
In the case that $\mstau{1} < \mst{1}$ also 
$\st_{1}\ra b \, \nu_{\tau} \, \stau_{1}$ may play a role.

In Fig. 7 a and b we show the parameter domains in the $M-\mu $
plane for the decays of $\wt t_1$ and $\wt b_1$, 
eqs. (\ref{stferm}), (\ref{sbferm}), (\ref{sqglu}), taking
$M_{\wt t_1} = 400$~GeV, $\tan \beta = 2$, and $M_{\wt b_1} = 400$~GeV, 
$\tan \beta = 30$, respectively. The parameter domains for the
$\wt \tau_1$ decays into neutralinos 
\noi
\begin{minipage}[t]{73mm}   % --- fig.7a ---
{\setlength{\unitlength}{1mm}
\begin{picture}(73,76)                        
%\put(0,0){\framebox(73,76){}}
\put(3,4){\mbox{\epsfig{figure=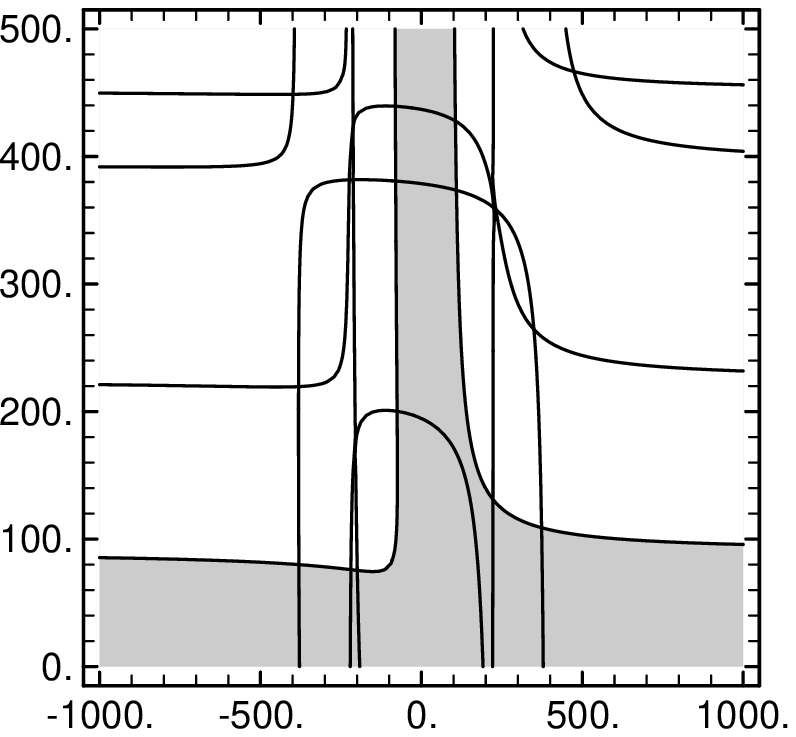,height=6.6cm}}}
\put(69.5,1){\makebox(0,0)[br]{{\small $\mu$~[GeV]}}}
\put(3,70){\makebox(0,0)[bl]{{\small $M$~[GeV]}}}
\put(20,58.5){{\small a}}
\put(60,59.5){{\small a}}
\put(20,51.5){{\small b}}
\put(60,53.0){{\small b}}
\put(20,34){{\small c}}
\put(60.5,35.5){{\small c}}
\put(35,56.0){{\small d}}
\put(35.0,51){{\small e}}
\put(35.2,31.2){{\small f}}
\end{picture}}
\refstepcounter{figure}    
%\label{fig:sl11prod}
\begin{small}
  {\bf Fig.~7a:}~Kinematically allowed parameter domains in the $(M,\,\mu)$
plane for $M_{\wt t_1} = 400$~GeV and $\tan \beta = 2$ for 
the decays:
a) $\wt t_1 \ra t \, \wt \chi^0_1$,
b) $\wt t_1 \ra b \, \wt \chi^+_1$, 
c) $\wt t_1 \ra t \, \wt \chi^0_2$, 
d) $\wt t_1 \ra t \, \wt \chi^0_3$, 
e) $\wt t_1 \ra b \, \wt \chi^+_2$, 
f) $\wt t_1 \ra t \, \wt \chi^0_4$. 
$\wt t_1 \ra c \, \wt \chi^0_1$ and
$\wt t_1 \ra b \, W^+ \, \wt \chi^0_1$ are allowed in the whole parameter 
range shown. The grey area is covered by LEP2 for $\sqrt{s} =$ 192 GeV.

\end{small} 
\end{minipage}
\hspace{3mm}
\begin{minipage}[t]{73mm}   % --- fig.7b ---
{\setlength{\unitlength}{1mm}
\begin{picture}(73,76)                        
%\put(0,0){\framebox(73,76){}}
\put(3,4){\mbox{\epsfig{figure=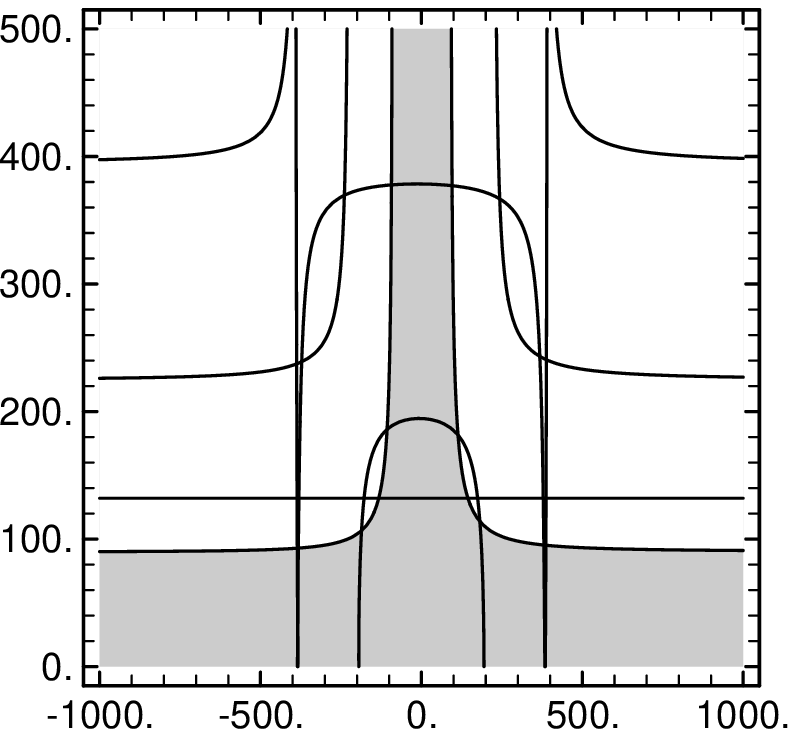,height=6.6cm}}}
\put(69.5,1){\makebox(0,0)[br]{{\small $\mu$~[GeV]}}}
\put(3,70){\makebox(0,0)[bl]{{\small $M$~[GeV]}}}
\put(20,53.5){{\small a}}
\put(60,53.5){{\small a}}
\put(20,34.5){{\small b}}
\put(60,34.5){{\small b}}
\put(30,54){{\small c}}
\put(47.5,54){{\small c}}
\put(35,50.0){{\small d}}
\put(35.5,30){{\small e}}
\put(20,24.5){{\small f}}
\end{picture}}
\refstepcounter{figure}    
%\label{fig:sl11prod}
\begin{small}
  {\bf Fig.~7b:}~Kinematically allowed parameter domains in the $(M,\,\mu)$
plane for $M_{\wt b_1} = 400$~GeV and $\tan \beta = 30$ for
the decays:
a) $\wt b_1 \ra b \, \wt \chi^0_2$, 
b) $\wt b_1 \ra t \, \wt \chi^-_1$, 
c) $\wt b_1 \ra b \, \wt \chi^0_3$, 
d) $\wt b_1 \ra b \, \wt \chi^0_4$, 
e) $\wt b_1 \ra t \, \wt \chi^-_2$, 
f) $\wt b_1 \ra b \, \wt g$. 
$\wt b_1 \ra b \, \wt \chi^0_1$  is  allowed in the whole parameter 
range shown. The grey area is covered by LEP2 for $\sqrt{s} =$ 192 GeV.
\end{small} 
\end{minipage}
\vspace{3mm}

\noi
are almost identical to those of
the corresponding $\wt b_1$ decays, if the masses of
 $\wt \tau_1$ and $\wt b_1$ are the same.

The branching ratios for the $\wt t_1$ decays as a function of
the mixing angle $\cos \Theta_{\wt t}$ are shown in Fig. 8a for 
$M_{\wt t_1} = 400$~GeV, $\tan \beta = 2$, and taking $M =
150$~GeV and $\mu = 500$~GeV. The decay into 
$b \wt \chi_1^+$ dominates near $\cos \Theta_{\wt t} = \pm 1$,
$\wt t_1 \approx \wt t_L$, whereas the decay into
$t \wt \chi_1^0$ dominates near  $\cos \Theta_{\wt t} = 0$,
$\wt t_1 \approx \wt t_R$. 
BR($\wt t_1 \ra b \wt \chi^+_1$) vanishes for $\cos \Theta_{\wt t}
\approx -0.3$ because gauge coupling and Yukawa coupling terms
cancel each other.
On the other hand, BR($\wt t_1 \ra t \wt \chi_1^0$) has a maximum for
$\cos \Theta_{\wt t} \approx -0.3$ because the two contributions add up.
Similarly, Fig. 8b exhibits the branching
ratios for the $\wt b_1$~decays
as a function of 
$\cos \Theta_{\wt b}$ for $M_{\wt b_1} = 400$~GeV, 
$\tan \beta = 30$, $M = 150$~GeV and $\mu = 500$~GeV.
Here the
branching ratio for the decay into $t \wt \chi^-_1$ is smaller
than that of $\wt t_1 \ra b \wt \chi^+_1$, because it has less
phase space. 
For $\tan \beta \grts 10$ the branching ratios are almost
symmetric under the simultaneous
interchange $\mu \leftrightarrow -\mu$ and 
$\cos \Theta_{\wt t} \leftrightarrow - \cos \Theta_{\wt t}$.
Note that in supergravity models \cite{Drees}, for large $\tan \beta$ and
large $| \mu |$, $\cos \Theta_{\sb}$ has the same sign as $\mu$, because
otherwise the parameter $A_b$ would be too large (see eq.(3)).

In Table~1 we list the most important
signatures for $\st_{1}$, $\sb_{1}$ and $\stau_{1}$
for $\sqrt{s} = 500$~GeV.
If the decays $\st_{1}\ra b \, \wt\chi^{+}_{1}$ or  
$\stau_{1}\ra \nu_{\tau} \, \wt\chi^{-}_{1}$ occur, the 
$\wt\chi^{\pm}_{1}$ would be discovered first and its properties
would be known. This would help identify these events.
The decay $\st_{1}\ra b\, W^{+}\, \wt\chi^{0}_{1}$ leads to the same
final states as $\st_{1}\ra b\, \wt\chi^{+}_{1}$ (provided 
$\wt \chi^{+}_{1} \ra H^+ \wt\chi^{0}_{1}$ is not allowed).
From the decay $\stau_{1}\ra \tau\, \wt\chi^{0}_{1}$ information about 
the neutralino parameters can be obtained by measuring the
$\tau$ polarization, as discussed in \cite{Nojiri}. 

\vspace{3mm}

\noi
\begin{minipage}[t]{73mm}   % --- fig.8a ---
{\setlength{\unitlength}{1mm}
\begin{picture}(73,76)                        
%\put(0,0){\framebox(73,76){}}
\put(3,4){\mbox{\epsfig{figure=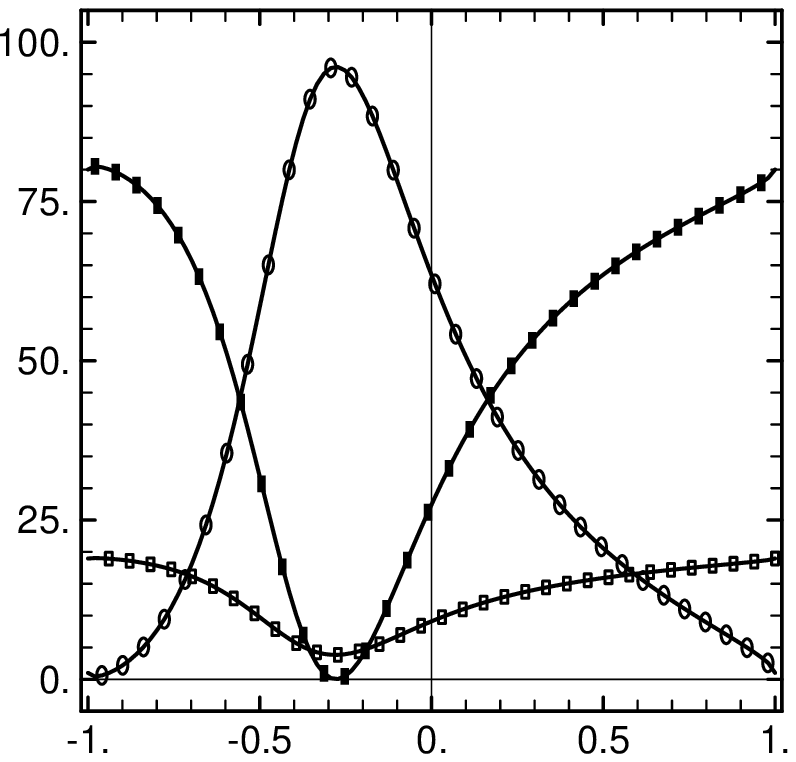,height=6.6cm}}}
\put(69.5,1){\makebox(0,0)[br]{{\small $\cos \Theta_{\wt t}$}}}
\put(3,70){\makebox(0,0)[bl]{{\small BR $(\st_{1})$~[\%]}}}
\end{picture}}
\refstepcounter{figure}    
%\label{fig:sl11prod}
\begin{small}
  {\bf Fig.~8a:}~Branching ratios for the $\wt t_1$ decays 
   as a function of the mixing angle $\cos \Theta_{\wt t}$ for 
   $M_{\wt t_1} = 400$~GeV, $\tan \beta = 2$, $M = 150$~GeV, 
   and $\mu = 500$~GeV.
   The curves correspond to the following transitions:
   $\circ \hspace{1mm} \wt t_1 \ra t \, \wt \chi^0_1$,
   \rechtl $\wt t_1 \ra t \, \wt \chi^0_2$,
   \recht $\wt t_1 \ra b \, \wt \chi^+_1$.
\end{small} 
\end{minipage}
\hspace{3mm}
\begin{minipage}[t]{73mm}   % --- fig.8b ---
{\setlength{\unitlength}{1mm}
\begin{picture}(73,76)
%\put(0,0){\framebox(73,76){}}
\put(3,4){\mbox{\epsfig{figure=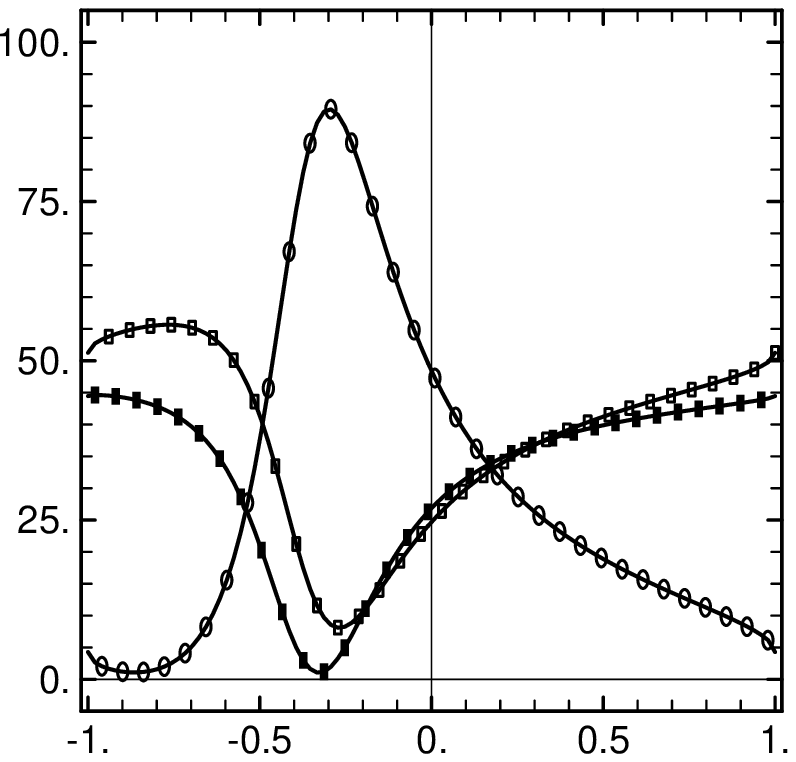,height=6.6cm}}}
\put(69.5,1){\makebox(0,0)[br]{{\small $\cos \Theta_{\wt b}$}}}
\put(3,70){\makebox(0,0)[bl]{{\small BR $(\sb_{1})$~[\%]}}}
\end{picture}}
\refstepcounter{figure}    
\begin{small}
  {\bf Fig.~8b:}~Branching ratios for the $\wt b_1$ decays 
   as a function of $\cos \Theta_{\wt b}$ for $M_{\wt b_1} = 400$~GeV, 
   $\tan \beta = 30$, $M = 150$~GeV, and $\mu = 500$~GeV. 
   The curves correspond to the following transitions:
   $\circ \hspace{1mm} \wt b_1 \ra b \, \wt \chi^0_1$,
   \rechtl $\wt b_1 \ra b \, \wt \chi^0_2$,
   \recht $\wt b_1 \ra t \, \wt \chi^-_1$.
\end{small} 
\end{minipage}
\vspace{3mm}

\begin{center}
\begin{tabular}{|l|l|}   
\hline 
   &  Signatures  \\
\hline  
  $\st_{1}\to b\,\chp_{1}$ & 1 $b$-jet + 1 $l^{+}$ + $\ptmiss$,
                             1 $b$-jet + 2 jets + $\ptmiss$ \\
\hline 
  $\st_{1}\to c\,\nt_{1}$  & 1 jet + $\ptmiss$ \\
\hline 
  $\sb_{1}\to b\,\nt_{1}$ & 1 $b$-jet + $\ptmiss$ \\
\hline
  $\sb_{1}\to b\,\nt_{2}$ & 1 $b$-jet + $l^{+}l^{-}$ + $\ptmiss$,
                            1 $b$-jet + 2 jets + $\ptmiss$ \\
\hline 
  $\stau_{1}\to \tau\,\nt_{1}$ & $\tau + \ptmiss$ \\
\hline
  $\stau_{1}\to \tau\,\nt_{2}$ & $\tau + l^{+}l^{-} + \ptmiss$,
                                 $\tau$ + 2 jets + $\ptmiss$ \\
\hline
  $\stau_{1}\to \nu_{\tau}\,\chm_{1}$ & $l^{-} + \ptmiss$,
                                        2 jets + $\ptmiss$ \\
\hline  
\end{tabular} 

\vspace{2mm}

\refstepcounter{table}   % set table counter +1 
\label{tab:signatures}
\noindent
{\bf Table~\arabic{table}:}~Expected signatures for 
$\st_{1}$, $\sb_{1}$, and $\stau_{1}$ production for
$\sqrt{s} =$~500~GeV. Due\\ \hspace*{-8mm} to pair production all combinations
of the corresponding signatures may occur.
\end{center}

%------------------------------------------------------------------------
\section{Stop Event Generation}
%------------------------------------------------------------------------

In this section we describe the event generator for $\eeto\st_1\bar{\st}_1$
with the stop decay modes $\st_1 \ra c \chiz$ and $\st_1 \ra b \chip$.
The chargino decays via $\chip \ra W^{+} \chiz$, where $W^{+}$
can be either virtual or real.
The event generator is based on the calculation
of the 4-momenta distributions of the stop and antistop decay products
\mbox{\chiz $c$ \chiz $\bar{c}$} and \mbox{\chip $b$ \chim 
$\bar{b}$}.
The large effects of QCD corrections are included in the
cross section calculation.
Stop production and decay have been defined as new processes in the 
PYTHIA program package~\cite{pythia}.
The event generation process includes the modelling of hadronic
final states. 

In the first step of the event generation, initial state photons are
emitted using the program package REMT~\cite{pythia} which takes into 
account the expected stop cross section from zero to the nominal 
center-of-mass energy. Beamstrahlung photons are emitted using the 
beam parameters of the NLC 1992 design.
The effective center-of-mass energy is calculated for the
initial production of the 4-momenta of the final-state particles.
These 4-momenta are then boosted to the lab-frame 
according to the momentum of the emitted photons.
For the hadronization process of the \cc\ in the 
\mbox{\chiz $c$ \chiz $\bar{c}$}
 and
of the \bb\ in the \mbox{\chip $b$ \chim $\bar{b}$} decay mode, 
a color string
with invariant mass of the quark-antiquark-system is defined. The possible
gluon emission and hadronization are performed using the Lund model
of string fragmentation with the PYTHIA program package~\cite{pythia}.
The Peterson {\it et \,al.}~\cite{peterson} fragmentation parameters 
for the $c$- and $b$-quarks are used: 
$\epsilon_c=0.03$ and $\epsilon_b=0.0035$.
Finally, short-lived particles decay into their observable final
state.
Details of the event generator and of a stop analysis at LEP2 energies
are given in~\cite{l3note}.

%------------------------------------------------------------------------
\section{Simulation and Selection}
%------------------------------------------------------------------------

\begin{figure}[ht]
\vspace*{-1.3cm}
\begin{center}
\mbox{\epsfig{file=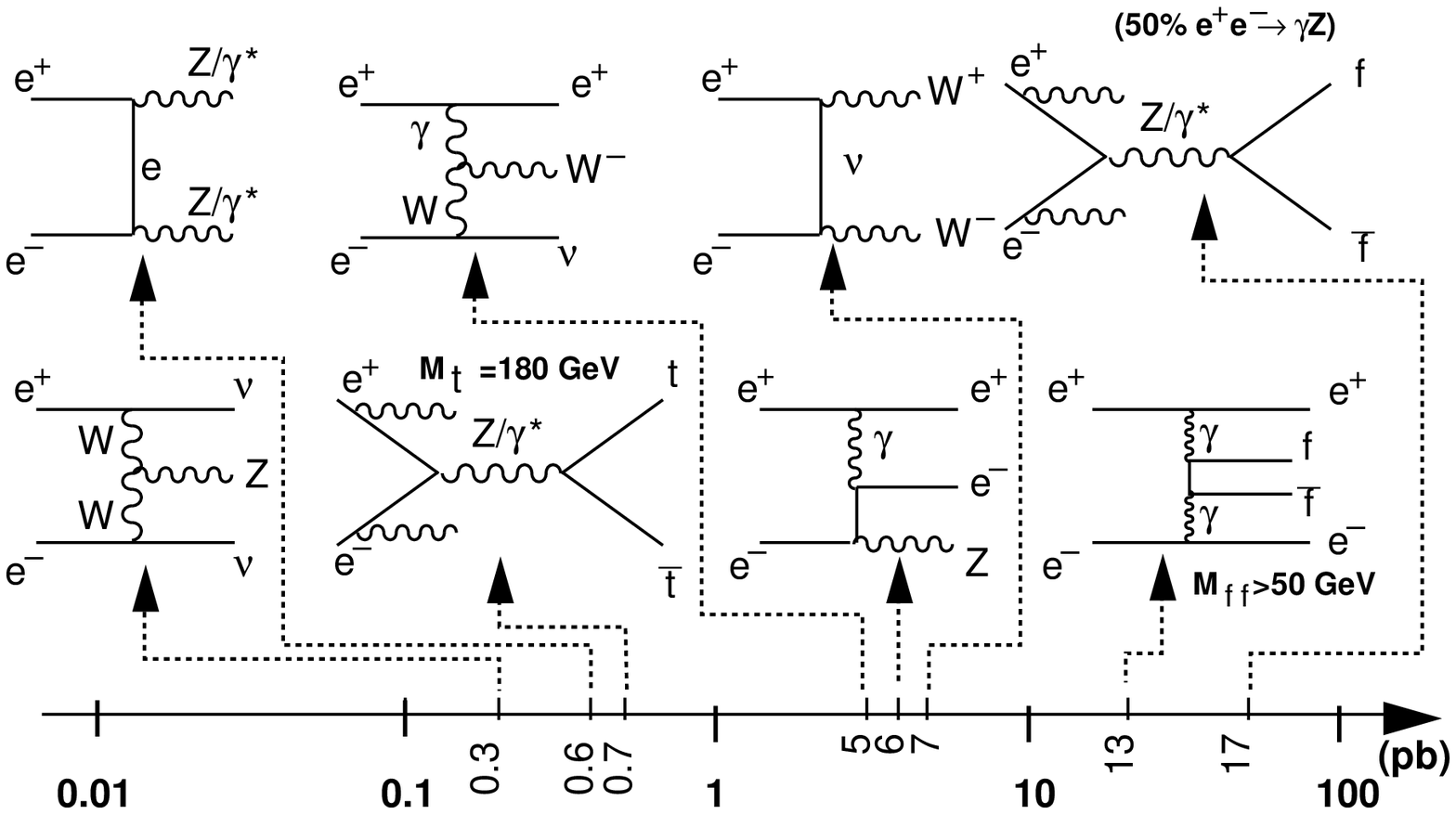,width=14cm}}
\end{center}
\vspace*{-0.8cm}
\caption{Background reactions and their cross sections for
        $\surd{s}=\gev{500}$.}                           
\label{fig:bg}
\vspace*{-0.2cm}
\end{figure}

The investigated background reactions and their cross sections
are shown in Fig.~\ref{fig:bg}. They are simulated for \lumifb{10},
and 1000 signal events are simulated in
the \chichi\ and \chipchip\ decay channels. 
The L3 detector at CERN including the upgrades for LEP2
served as an example for an \ee\ \gev{500} detector. Details of the
parametric detector simulation are given in~\cite{zphys}.
An important feature is the overall hadronic energy resolution of
about 7\%. 

In both channels, 
the \chiz 's escape the detector and cause large missing energy.
In the case \chichi, the $c$-quarks form mostly two acoplanar jets. 
A mass combination of $M_{\st_1}=\gev{180}$ and $m_{\nt_1}=\gev{100}$
is investigated in detail. For {\chipchip} on average the
visible energy is larger. In this channel, the mass combination
$M_{\st_1}=\gev{180}$, 
$m_{\chp_1}=\gev{150}$, and $m_{\nt_1}=\gev{60}$ has been studied.
Typically four jets are formed, two from the $b$-quarks,
and two from the boosted $W$'s.

In the first step of the event selection, unbalanced hadronic events
are selected using the following selection requirements:
$$
25 <  {\mathrm{hadronic~clusters}} < 110,~~~~~
0.2 < E_{vis}/\sqrt{s} < 0.7,
$$
$$
E^{\mathrm{imb}}_{\parallel}/E_{\mathrm{vis}} < 0.5,~~~~~
\mathrm{Thrust} < 0.95,~~~~~
\abs{\cos\theta_{\mathrm{Thrust}}}<0.7~.
$$

\begin{table}[h]
\vspace*{-0.4cm}
\renewcommand{\arraystretch}{1.1}
\begin{center}
\begin{tabular}{|c|c|c|c|c|c|c|c|c|}\hline
Channel &\chichi&\chipchip& qq & WW & eW$\nu$& tt &ZZ  & eeZ \\ \hline
Total (in 1000)&1& 1      &125 & 70 &  50    &  7 & 6  & 60    \\ \hline
After preselection (in 1000) & 0.4   & 0.7   & 1.7& 2.2&  3.2   & 1.3& 0.2& 0.3   \\ \hline
\end{tabular}
\end{center}
\vspace*{-0.5cm}
\caption{\label{tab:pres} Expected events per 10~fb$^{-1}$ at 
$\surd{s} = \gev{500}$, and number of events after the preselection as 
defined in the text.}
\renewcommand{\arraystretch}{1.0}
\end{table}

\noi
A large part of the background of back-to-back events without missing
energy is rejected.
Table~\ref{tab:pres} shows the number of initially produced events per
${\cal L}$ = 10 $\rm fb^{-1}$ at $\sqrt{s}=500$~GeV, and the number 
of events which pass this preselection. 
The requirement of a large number of hadronic clusters
removes \ee, \mm, and most \tautau\ events. The minimum energy cut reduces
most of the $\gamma\gamma$ events and ensures almost
100\% trigger efficiency.
The background from $\gamma\gamma$ events can, in addition, be strongly 
reduced by rejecting events where a scattered initial electron is detected
at low angles.
The upper energy cut reduces all standard background reactions.
Beam gas events and events where much energy goes undetected along the beam
axis are removed by rejection of events with very large
parallel imbalance. 
The thrust cut removes remaining \tautau\ events and reduces largely \qq\ and \ZZ\ background.
The $\cos\theta_{\mathrm{Thrust}}$ cut removes events where most 
probably much energy escaped 
undetected along the beam axis.

\begin{figure}[t]
\vspace*{-0.7cm}
\begin{tabular}{p{8cm} p{8cm} }
\mbox{\epsfig{file=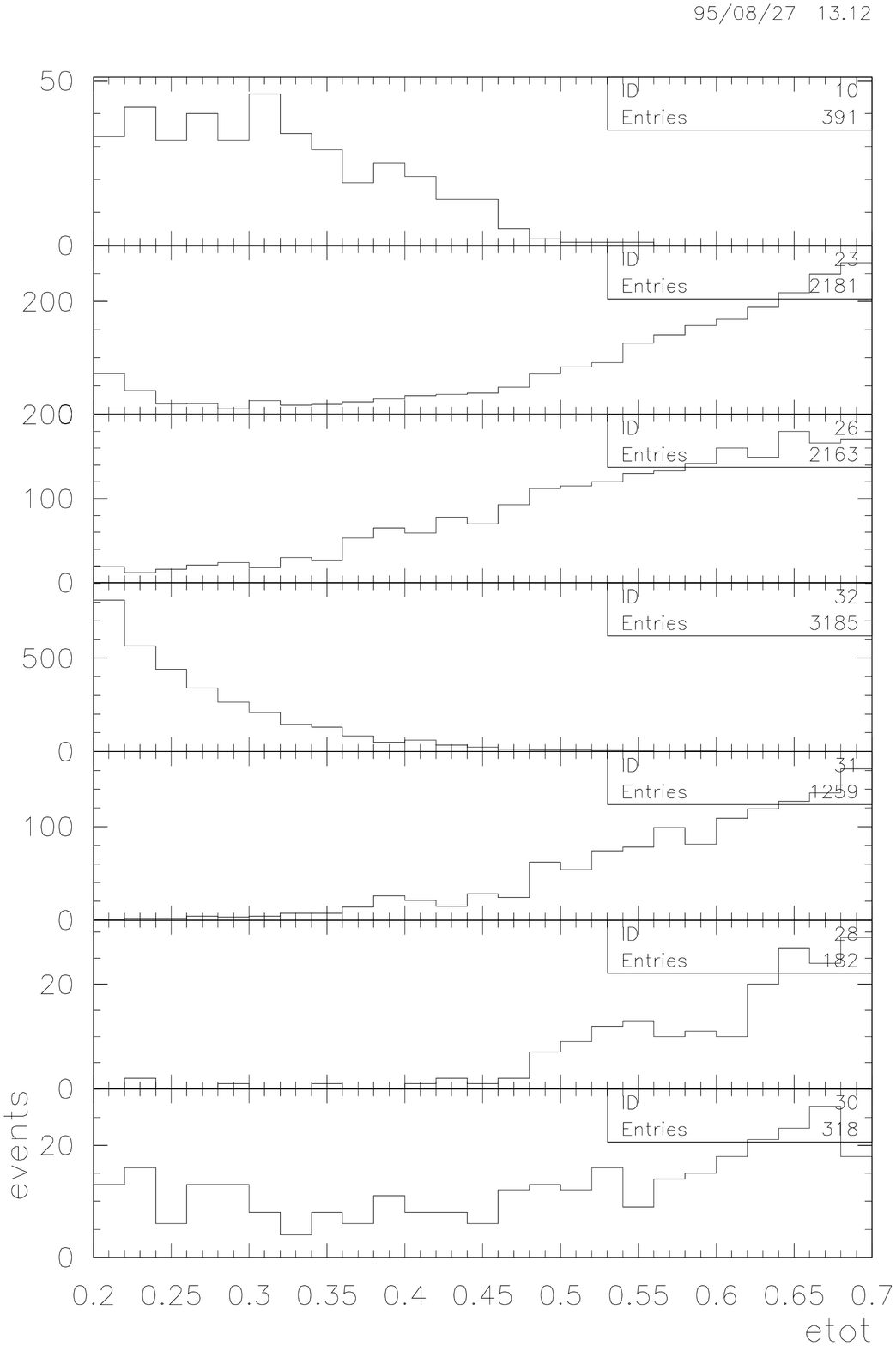,width=7.5cm,
       bbllx=0pt,bblly=55pt,bburx=560pt,bbury=800pt,clip=}}%
\mbox{\epsfig{file=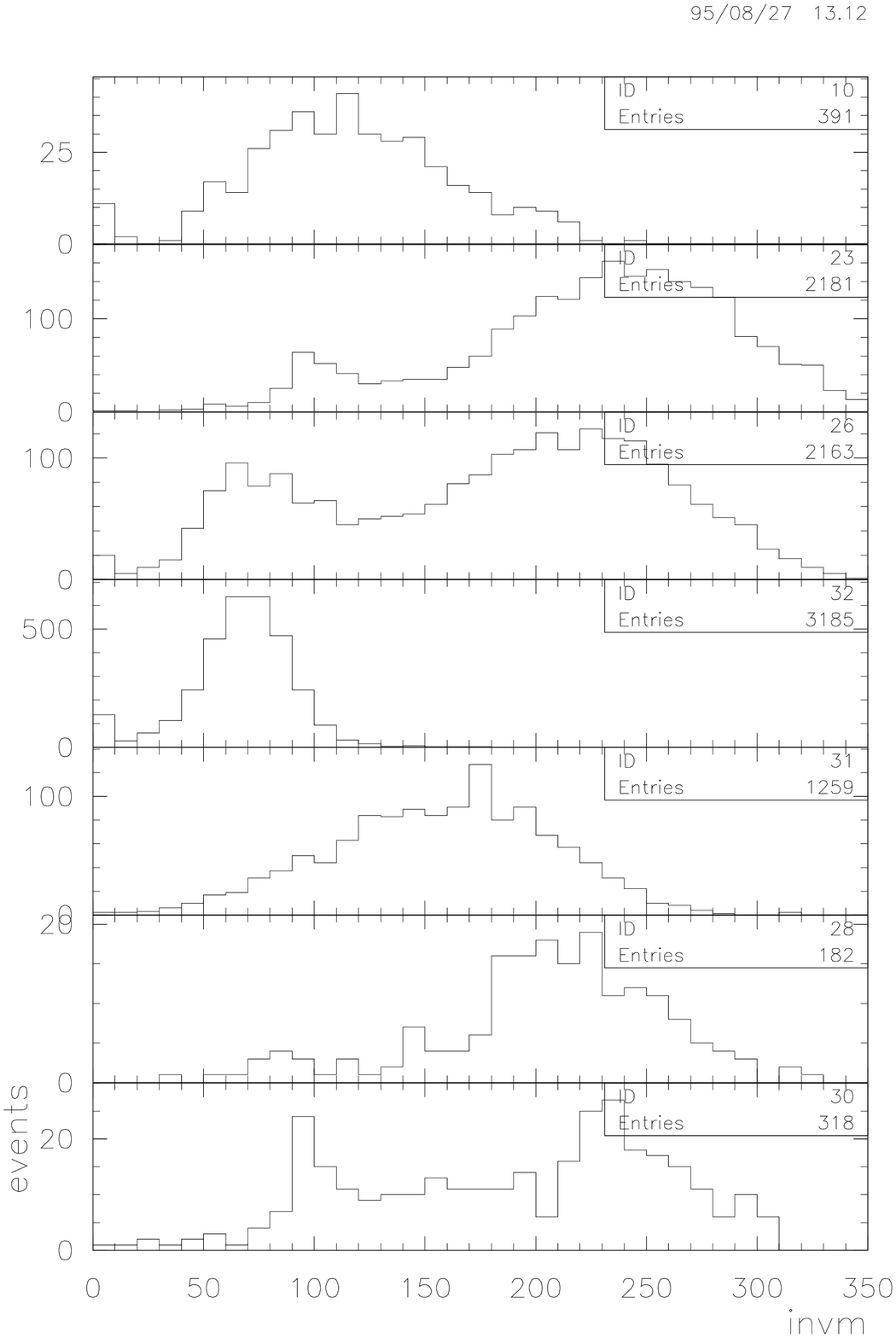,width=7.5cm,
       bbllx=0pt,bblly=55pt,bburx=560pt,bbury=800pt,clip=}}%
\end{tabular}
\begin{tabular}{p{0.1cm} p{7cm} p{-0.5cm} p{6.5cm} }
     &
\vspace*{-0.8cm}
\caption{\label{fig:evis} $E_{vis}/\surd s<0.4$ for 
         \chichi, qq, WW, eW$\nu$, tt, ZZ, eeZ.} 
&
     &
\vspace*{-0.8cm}
\caption{\label{fig:minv} $m_{\mathrm{inv}}>\gev{120}$
                for \chichi, qq, WW, eW$\nu$, tt, ZZ, eeZ.}
\end{tabular}
\begin{tabular}{p{8cm} p{8cm} }
\mbox{\epsfig{file=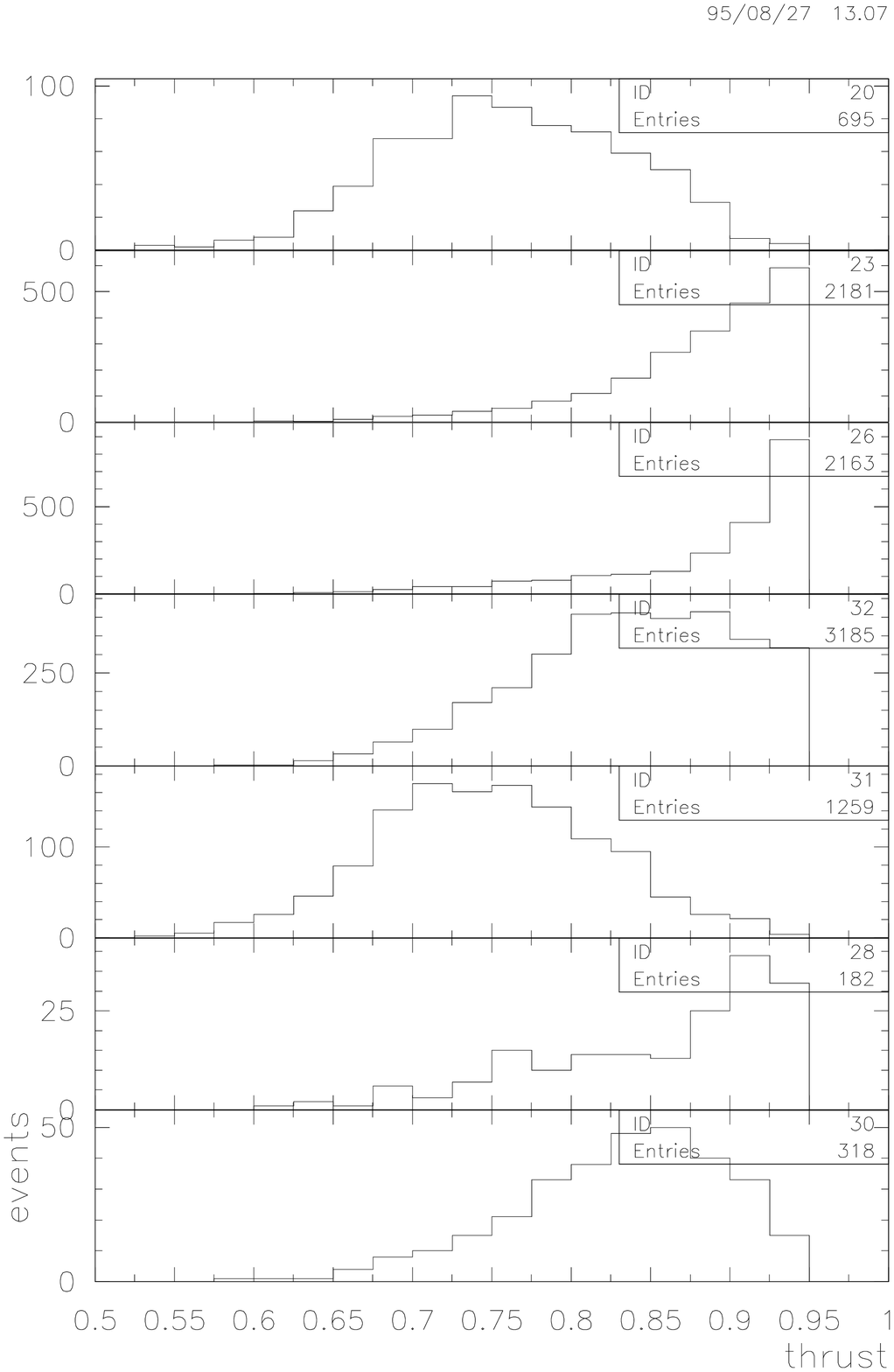,width=7.5cm,
       bbllx=0pt,bblly=55pt,bburx=560pt,bbury=800pt,clip=}}%
\mbox{\epsfig{file=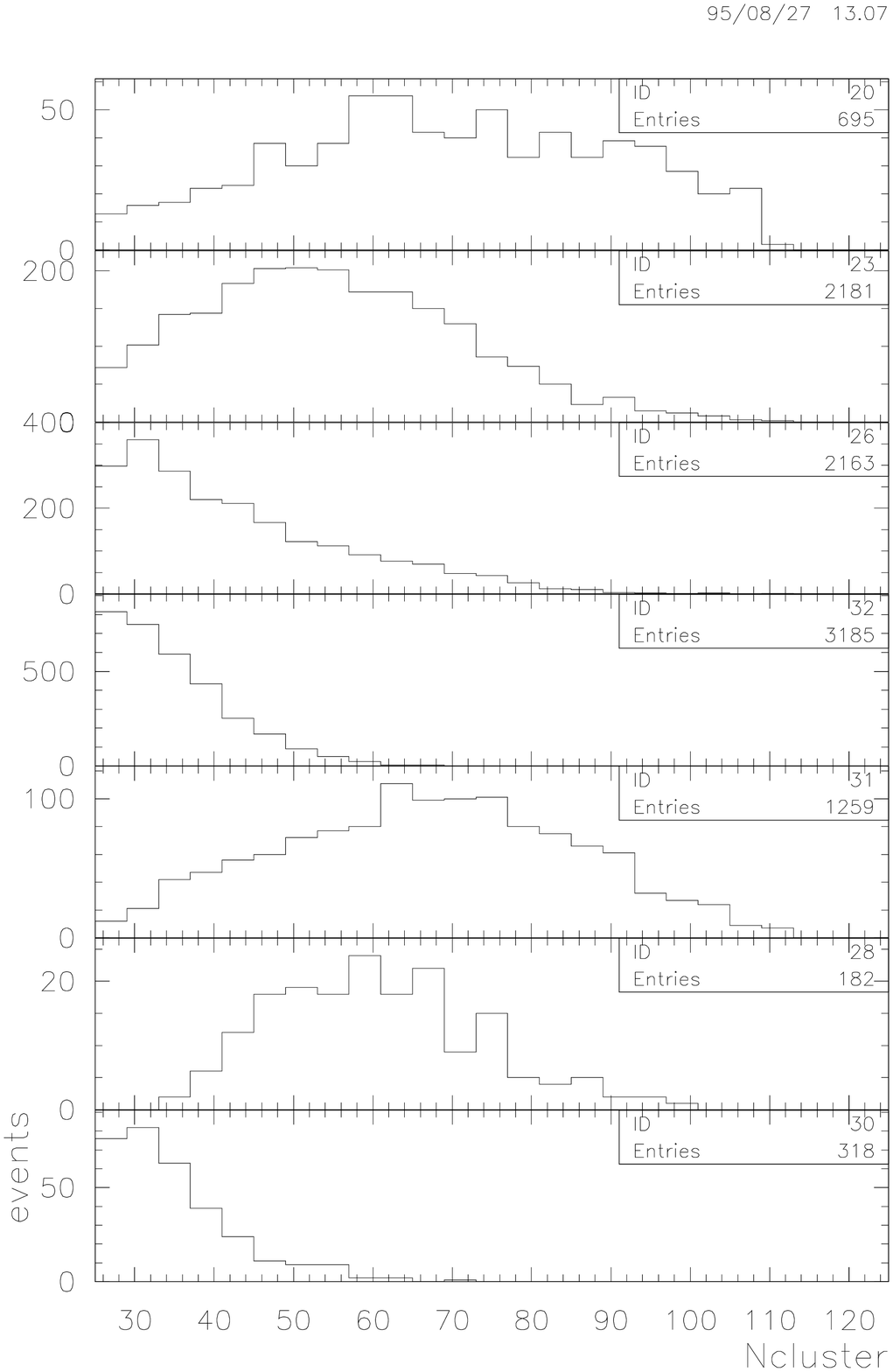,width=7.5cm,
       bbllx=0pt,bblly=55pt,bburx=560pt,bbury=800pt,clip=}}%
\end{tabular}
\begin{tabular}{p{0.1cm} p{7cm} p{-0.5cm} p{6.5cm} }
     &
\vspace*{-0.8cm}
\caption{\label{fig:thrust} Thrust $<0.85$ 
         for \chipchip, qq, WW, eW$\nu$, tt, ZZ, eeZ.} &
     &
\vspace*{-0.8cm}
\caption{\label{fig:cluster} Ncluster $\geq 60$
         for \chipchip, qq, WW, eW$\nu$, tt, ZZ, eeZ.}
\end{tabular}
\end{figure}

The final \chichi\ event selection is summarized in Table~\ref{tab:final}.
The following cuts are applied: 
\begin{itemize}
\vspace*{-3mm}
\item A hard upper energy cut reduces all standard background
      except $e W \nu$  (Fig.~\ref{fig:evis}).
\vspace*{-8mm}
\item Jets are clustered using the JADE algorithm.
      The y-cut value is optimized to obtain two jets for the signal.
\vspace*{-3mm}
\item Semileptonic decays of the top quark can induce missing energy.
      These events are partly removed by requiring no isolated electron 
      or muon.
\vspace*{-3mm}
\item Events with large longitudinal energy imbalance are removed 
      where probably much energy escaped undetected along the beam axis.
\vspace*{-3mm}
\item The invariant mass of the two jets is required to be larger
      than \gev{120} to remove almost entirely $eW\nu$ events
      (Fig.~\ref{fig:minv}).
\vspace*{-3mm}
\item The acoplanarity angle is defined as the angle between 
      the jets in the plane perpendicular to the beam axis. 
      A maximum value of 2.9~rad is important to reduce the remaining 
      background. \vspace*{-2mm}
\end{itemize}

\begin{figure}[t]
\vspace*{-0.3cm}
\begin{tabular}{p{0.48\linewidth}p{0.48\linewidth}}
\mbox{\epsfig{file=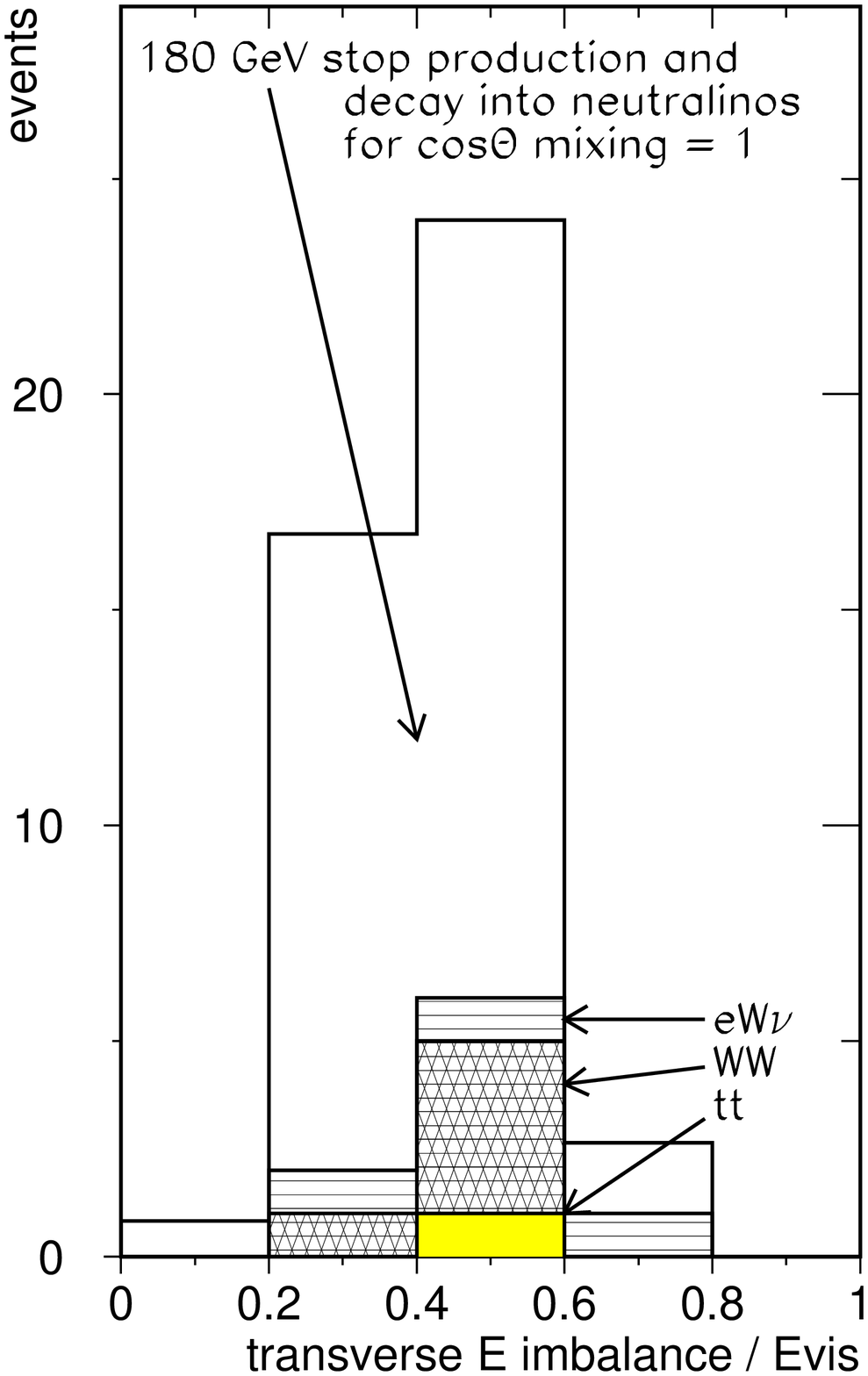,height=10.9cm}} &
\mbox{\epsfig{file=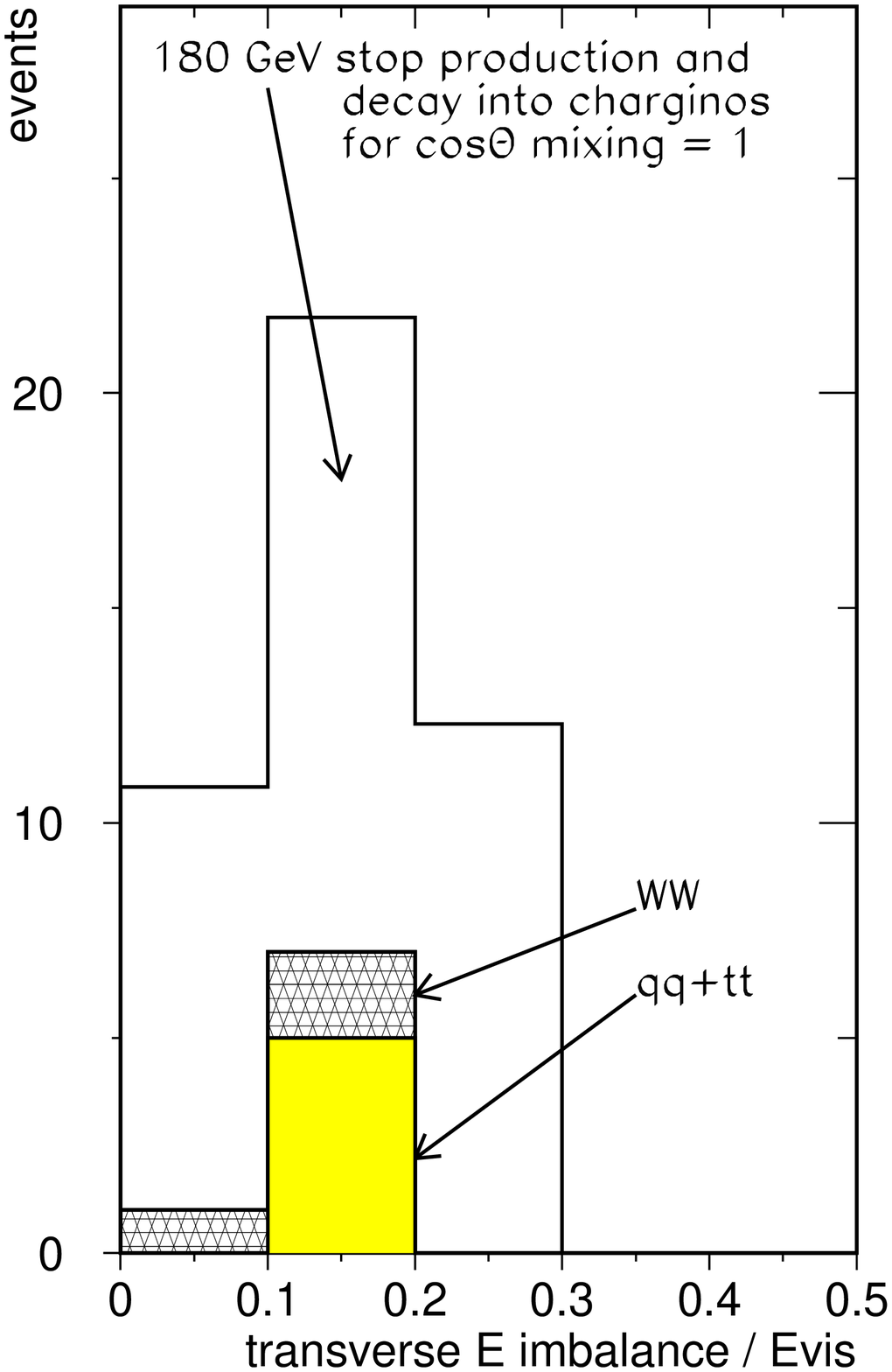,height=10.9cm}}
\end{tabular}
\vspace*{-1.3cm}
\caption{\label{fig:ee500tbtb} \baselineskip=12pt
Left: Sensitivity for an 
$\eeto\st_1\bar{\st}_1 \rightarrow \nt_1 c\nt_1\bar{c}$ 
signal.
Open histograms show the simulated signal, solid
and hatched histograms show the remaining background
after all selection cuts are applied.
Right:  Sensitivity for an
$\eeto\st_1\bar{\st}_1 \rightarrow \chp_1 b\chm_1\bar{b}$ 
signal.
 }
\end{figure}

The result of this study is
4.3\% detection efficiency and 9 background events.
A detection confidence level of 3$\sigma$ (99.73\%) is expected 
for a cross section of 23~fb. Expected signal and background
are shown in Fig.~\ref{fig:ee500tbtb}.

\begin{table}[hb]
\renewcommand{\arraystretch}{1.1}
\vspace*{-0.8cm}
\begin{center}
\begin{tabular}{|c|c|c|c|c|c|c|c|}\hline
Channel &$\nt_1$c$\nt_1 \bar{c}$& qq & WW &eW$\nu$& tt &ZZ  & eeZ   \\ \hline
Total (in 1000)    & 1     &125 & 70 &  50 &  7 & 6  & 60    \\ \hline

After Preselection & 391   &1652&2163&3185 &1259&182 & 318   \\ \hline

$E_{vis}/\sqrt{s}<0.4$  
                  & 332   &{\bf 202}&{\bf 285}&3032   &{\bf 70}&{\bf 4}&{\bf 98} \\ \hline

Njet $=2$          & 293   & 172& 182&2892   &{\bf 17}&  3 &  72 \\ \hline

No isolated e or $\mu$
                   & 218   & 152&{\bf 98}&2757&{\bf 5}&  3 &{\bf 9} \\ \hline

$E^{imb}_{\parallel}/E_{vis}<0.3$
                   &185 &{\bf 101}&  70&2049   &   5&  2 &   4   \\ \hline

Invariant\,mass\,of\,jets$>$120GeV
                   & 52    &{\bf 25}&{\bf 12}&{\bf 7}&1&  0 &   0 \\ \hline
Acoplanarity $<2.9$rad&43     &{\bf 0} &{\bf 5}&{\bf 3}&  1&  0 &   0  \\ \hline
\end{tabular}
\end{center}
\vspace*{-0.6cm}
\caption{\label{tab:final} Final event selection cuts,
         expected signal efficiencies, and the number of
         expected background events. Bold face numbers indicate major 
         background reductions.}
\vspace*{-2mm}
\renewcommand{\arraystretch}{1.0}
\end{table}

The final \chipchip\ event selection is summarized in Table~\ref{tab:final2}.
Here the cuts are:
\begin{itemize}
\sloppy
\vspace*{-3mm}
\item A hard lower energy cut reduces most of the $eW\nu$ background.
\vspace*{-3mm}
\item Topologies with back-to-back jets are reduced by an upper cut on
      the event thrust (Fig.~\ref{fig:thrust}).\vspace*{-3mm}
\item A lower cut on the number of hadronic clusters reduces efficiently
      low-multiplicity background final states
      (Fig.~\ref{fig:cluster}).\vspace*{-3mm}
\item Jets are clustered using the JADE algorithm.
      The y-cut value is optimized to obtain four jets for the signal.
\vspace*{-3mm}
\item Events with an isolated electron or muon are rejected.
\vspace*{-3mm}
\item An upper cut on the visible energy reduces \qq, \WW, and 
      $t \bar{t}$ background.
\vspace*{-3mm}
\item Finally, the remaining $t \bar{t}$ background events are 
reduced by requiring less than 30\% perpendicular energy 
imbalance.\vspace*{-3mm}
\end{itemize}

Concerning the number of $b$-quarks per event, the decay
$\chipchip \ra W^+\chiz bW^-\chiz \bar{b}$ leads to the same final 
states as expected for $t \bar{t}$ background. 
Therefore, the tagging of $b$-quarks has not 
proved to be efficient to reduce this background.

The result of this study is
4.5\% detection efficiency and 8 background events.
A detection confidence level of 3$\sigma$ (99.73\%) is expected 
for a cross section of 19~fb.
Expected signal and background are shown in 
Fig.~\ref{fig:ee500tbtb}.

\clearpage

\clearpage
\nopagebreak
\begin{figure}[hb]
\vspace*{-0.2cm}
\begin{tabular}{p{0.48\linewidth}p{0.48\linewidth}}
\mbox{\epsfig{file=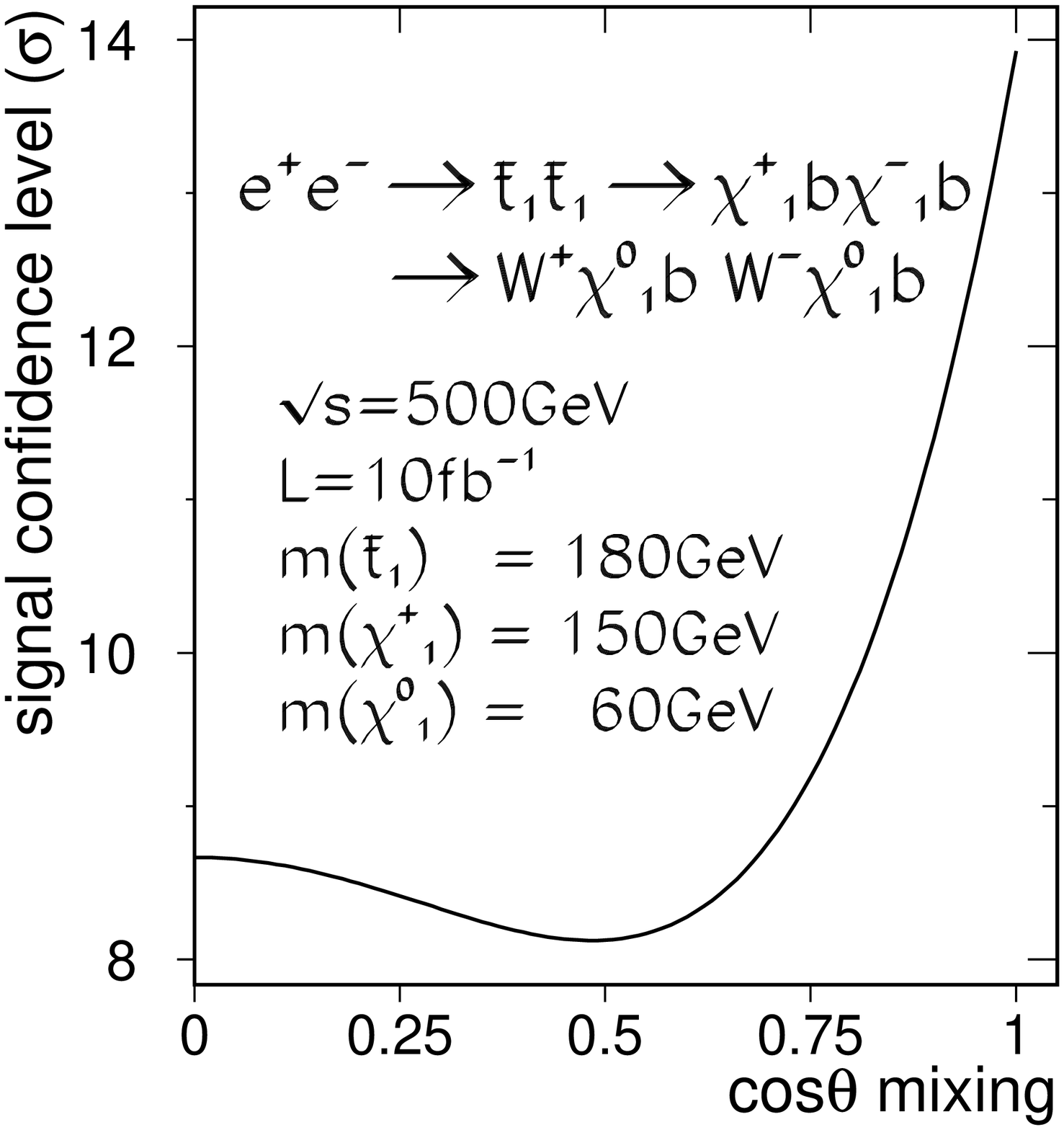,height=8cm}} &
\mbox{\epsfig{file=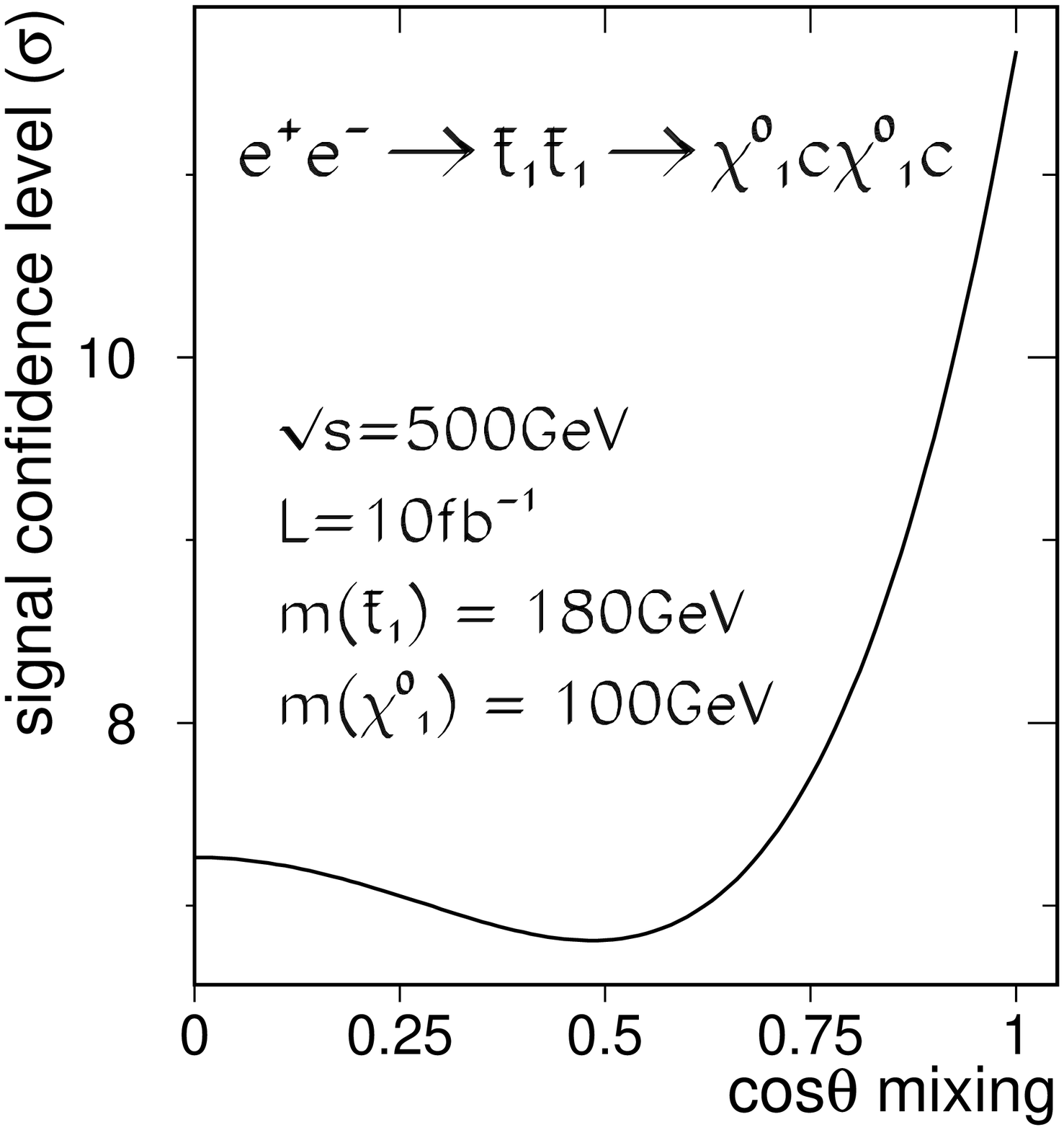,height=8cm}}
\end{tabular}
\vspace*{-1.2cm}
\caption{\label{fig:conf}  Detection confidence levels.
Left: \chipchip\ channel. Right: \chichi\ channel.}
\vspace*{-0.1cm}
\end{figure}

\clearpage
\begin{table}[ht]
\renewcommand{\arraystretch}{1.05}
\vspace*{-0.7cm}
\begin{center}
\begin{tabular}{|c|c|c|c|c|c|c|c|}\hline
Channel            &$\chp_1 b\chm_1\bar{b}$& qq & WW &eW$\nu$& tt &ZZ  & eeZ   \\ \hline
Total (in 1000)    & 1    &125 & 70 &  50   &  7 & 6  & 60    \\ \hline

After Preselection & 695  &1652&2163&3185   &1259&182 & 318   \\ \hline

$E_{vis}/\sqrt{s}>0.35$ 
                   & 610  &1494&2011&{\bf 337}&1234&178 & 239 \\ \hline

Thrust $<0.85$     & 536  &{\bf 326}&{\bf 420}&{\bf 24}&1141&{\bf 69}&137 \\ \hline   

Ncluster $\geq 60$ & 399  & 195&{\bf 134}&{\bf 0}  & 769& 41 &{\bf 3}     \\ \hline 

Njet $=4$          & 211  &{\bf 53}&{\bf 72}& 0  & 432& 22 &   0    \\ \hline

No isolated e or $\mu$& 99&  41&  49&   0  &{\bf 105}& 16 & 0 \\ \hline

$E_{vis}/\sqrt{s}<0.55$&57&{\bf 3}&{\bf 8}& 0 &{\bf  23}& 0 & 0 \\ \hline

$E^{imb}_{\perp}/E_{vis}<0.3$& 45     &  1& 3& 0 & {\bf 4}&  0 & 0 \\ \hline

\end{tabular}
\end{center}
\renewcommand{\arraystretch}{1.0}
\vspace*{-0.6cm}
\caption{\label{tab:final2} Final event selection cuts, 
         expected signal efficiencies, and the number of 
         expected background events. Bold face numbers indicate major
         background reductions.}
\vspace*{-0.1cm}
\end{table}

At a future \ee\ collider with $\sqrts=\gev{500}$, a large discovery
potential for scalar top quarks is already expected within one
year of data-taking (\lumifb{10}).
Detector performances known from LEP detectors result in good
background reduction. Full hermeticity of the detector is essential.

The confidence level for discovering a signal is shown in
Fig.~\ref{fig:conf}, where the confidence level is given in
$\sigma = N_{\mathrm{expected}} / \sqrt{N_{\mathrm{background}}}$.
The sensitivity is sufficient to discover a \gev{200} stop 
independently of the values of the mixing angle
with $3\sigma$ in both \chiz $c$ and \chip $b$ decay modes 
for the investigated neutralino and chargino mass
combinations. A complete set of mass combinations remains to be
studied. Beam polarization could be crucial for determining 
the stop mass after a discovery.

%------------------------------------------------------------------------
\section{Summary}
%------------------------------------------------------------------------

In this contribution we have discussed the production of stop, sbottom,
and stau pairs in $e^+e^-$ annihilation in the energy range
$\sqrt{s} = 500$~GeV to $2$~TeV.
We have presented numerical predictions within the Minimal 
Supersymmetric Standard Model for the production cross sections 
and the decay rates and analyzed their SUSY parameter dependence.
If $\tan \beta \grts 10$, not only the $t$ Yukawa terms,
but also the $b$ and $\tau $ Yukawa terms have important effects. 
The production 
cross sections as well as the decay rates of stops, sbottoms
and staus depend in a characteristic way on the mixing angles.

A Monte Carlo study of $\eeto\stst$ at $\sqrt{s} = 500$~GeV with 
the decays $\st_{1}\ra c\, \wt\chi^{0}_{1}$ and 
$\st_{1}\ra b\, \wt\chi^{+}_{1}$ has been performed for $\mst{1} = 180$~GeV,
$\mnt{1} = 100$~GeV, and $\mst{1} = 180$~GeV,
$\mch{1} = 150$~GeV, $\mnt{1} = 60$~GeV, respectively. 
A suitable set of kinematical cuts has been applied to reduce the known
background reactions. Detection confidence levels as a funtion of 
$\cos\Theta_{\st}$ have been given. 
In summary, an \ee ~collider is an ideal machine for 
detecting and studying scalar top and bottom quarks and scalar tau 
leptons.

%------------------------------------------------------------------------
\section*{Acknowledgements}
%------------------------------------------------------------------------

We thank our colleagues at this Workshop for many useful discussions.
This work was supported by the
``Fonds zur F\"orderung der wissenschaftlichen
Forschung'' of Austria, project no. P10843-PHY.

%------------------------------------------------------------------------


\begin{thebibliography}{99}
%------------------------------------------------------------------------

\bibitem{Ellis} J. Ellis, S. Rudaz, Phys. Lett. B128 (1983) 248

\bibitem{Altarelli} G. Altarelli, R. R\"uckl, Phys. Lett. B144 (1984) 126\\
I. Bigi, S. Rudaz, Phys. Lett. B153 (1985) 335

\bibitem{Boer} 
W. de Boer, R. Ehret, D. I. Kazakov, Phys. Lett. B334 (1994) 220 \\
W. de Boer et al., these proceedings           

\bibitem{Drees} See for a review e.g., M. Drees and S.P. Martin,
Wisconsin preprint, MADPH-95-879

\bibitem{Bartl94} A. Bartl, W. Majerotto, W. Porod, Z. Phys. C64 (1994) 499

\bibitem{Kon}  M. Fukugita, H. Murayama, M. Yamaguchi, T.
Yanagida, Phys. Rev. Lett. 72 (1994) 3009 \\
T. Kon, T. Nonaka, preprint ITP-SU-94/02, RUP-94-06 \\
J. D. Wells, G. L. Kane, Phys. Rev. Lett. 76 (1996) 869\\
A. Brignole, F. Feruglio, F. Zwirner, CERN-TH/95-340

\bibitem{Haber} H.E. Haber, G.L. Kane, Phys. Rep. 117 (1985) 75

\bibitem{Gunion} J. F. Gunion, H. E. Haber, Nucl. Phys. B272 (1986) 1

\bibitem{Grivaz} J.-F. Grivaz, Rapporteur Talk, International Europhysics
Conference on High Energy Physics, Brussels, 1995

\bibitem{Opal94} OPAL Collaboration, R. Akers et~al.,
Phys. Lett. B337 (1994) 207 \\
ALEPH Collaboration, Contribution \#0416 to the International Europhysics
Conference on High Energy Physics, Brussels, 1995

\bibitem{Nowak}
ALEPH Collaboration, CERN-PPE/96-10, Jan. 1996 \\
H. Nowak and A. Sopczak, L3 Note 1887, Jan. 1996 \\
S. Asai and S. Komamiya, OPAL Physics Note PN-205, Feb. 1996

\bibitem{D0} D0 Collaboration, FERMILAB-Conf-95/393-E, Proc. of the
10th Topcial Workshop on ``Proton--Antiproton Collider Physics'',
FNAL (1995)

\bibitem{Drees90} M. Drees, K. I. Hikasa, Phys. Lett. B252 (1990) 127

\bibitem{Bartl94a} A. Bartl, H. Eberl, W. Majerotto, W. Porod,
Proc. of the US-Polish Workshop ``Physics from Planck Scale to
Electroweak Scale'', Warshsaw 1994, p. 370, World Scientific
(P. Nath, T. Taylor, S. Pokorski eds.)

\bibitem{Hikasa} K. I. Hikasa, M. Kobayashi, Phys. Rev. D36 (1987) 724

\bibitem{Bartl96} A. Bartl, H. Eberl, S. Kraml, W. Majerotto, W. Porod,
preprint UWThPh-1996-18, HEPHY-PUB 642/96

\bibitem{Eberl} H. Eberl, A. Bartl, W. Majerotto,
preprint UWThPh-1996-6, HEPHY-PUB 640/96

\bibitem{Beenaker} W. Beenakker, R. H\"opker, P. M. Zerwas,
Phys. Lett. B349 (1995) 463

\bibitem{Peskin} See e.g. M. Peskin, 17th SLAC Summer Institute,
SLAC-PUB-5210 (1990)

\bibitem{Bartl91}
A. Bartl, W. Majerotto, B. M\"osslacher, N. Oshimo, S. Stippel,
Phys. Rev. D43 (1991) 2214

\bibitem{Nojiri} M. Nojiri, Phys. Rev. D51 (1995) 6281

% --- Spoczak's references ---

\bibitem{pythia} T.~Sj{\"o}strand, Comp. Phys. Comm. 82 (1994) 74

\bibitem{peterson} C.~Peterson \etal, Phys. Rev. D27 (1983) 105

\bibitem{l3note} A.~Sopczak, L3 note \#1860 (1995), to be published in the
LEP2 CERN Workshop report

\bibitem{zphys}A.~Sopczak,
Proc. Workshop on physics and experiments with linear \ee\ colliders,
Waikoloa, Hawaii, USA, 26--30 April 1993 (World Scientific) p. 666;
Z. Phys. C65 (1995) 449

\end{thebibliography}
\end{document}